\documentclass[aps,pre,reprint,showpacs,floatfix,superscriptaddress]{revtex4-1}
\usepackage[colorlinks,linkcolor=red,anchorcolor=red,citecolor=blue,urlcolor=blue]{hyperref}
\usepackage{amsmath,txfonts,sidecap}
\usepackage{graphicx}
\usepackage{dcolumn,enumitem}
\usepackage{float}
\usepackage{subfig}
\usepackage{multirow}
\allowdisplaybreaks[3]

\begin{document}

\title{Temporal Link Prediction in Social Networks Based on Agent Behavior Synchrony and a Cognitive Mechanism}

\author{Yueran Duan}
\email{sarah\_623@outlook.com}
\affiliation{School of Information Engineering, China University of Geosciences, Beijing, China}
\author{Mateusz Nurek}
\email{mateusz.nurek@pwr.edu.pl}
\affiliation{Department of Artificial Intelligence, Wrocław University of Science and Technology, Wrocław, Poland}
\author{Qing Guan}
\email{guanqing35@126.com}
\author{Radosław Michalski}
\email{radoslaw.michalski@pwr.edu.pl}
\affiliation{Department of Artificial Intelligence, Wrocław University of Science and Technology, Wrocław, Poland}
\author{Petter Holme}
\email{petter.holme@aalto.fi}
\affiliation{Department of Computer Science, Aalto University, Finland}
\affiliation{The Center for Computational Social Science, Kobe University, Japan}

\begin{abstract}
Temporality, a crucial characteristic in the formation of social relationships, was used to quantify the long-term time effects of networks for link prediction models, ignoring the heterogeneity of time effects on different time scales. In this work, we propose a novel approach to link prediction in temporal networks, extending existing methods with a cognitive mechanism that captures the dynamics of the interactions.  Our approach computes the weight of the edges and their change over time, similar to memory traces in the human brain, by simulating the process of forgetting and strengthening connections depending on the intensity of interactions. We utilized five ground-truth datasets, which were used to predict social ties, missing events, and potential links. We found: (a) the cognitive mechanism enables more accurate capture of the heterogeneity of the temporal effect, leading to an average precision improvement of 9\% compared to baselines with competitive AUC.  (b)  the local structure and synchronous agent behavior contribute differently to different types of datasets. (c) appropriately increasing the time intervals, which may reduce the negative impact from noise when dividing time windows to calculate the behavioral synchrony of agents, is effective for link prediction tasks.
\end{abstract}

\maketitle

\section{Introduction}
\label{introduction}

Link prediction in social networks provides a means to address data incompleteness, unveiling latent relationships or predicting the future evolution of network structure~\cite{lu_link_2011}. With the development of complex social network modeling, link prediction is rapidly becoming adopted as a research method in the social sciences~\cite{daud_applications_2020}. Temporality, as a crucial feature of real social networks, makes interactions within social networks evolve over time, implying not only the emergence and disappearance of links but also a shift in perception regarding their significance~\cite{michalski2021social,flamino2022modeling,michalski_temporal_2022,yanchenko_rev}. Some models attempt to account for the temporal dimension by incorporating time-decay functions to represent the forgetting effect~\cite{liao_ranking_2017,butun_extension_2018,liao_temporal_2019}. Most of these models regard social agents as isolated individuals, ignoring the network effects generated by synchronous interaction between nodes. 

Previous studies have shown that agent behavioral synchrony plays an important role in revealing the structure of social networks~\cite{pfaus_distal_2023}. Some scholars have described inter-agent behavioral synchrony through mutual information~\cite{jiang_link_2023} or time vector similarity~\cite{duan2023temporal} and developed temporal link prediction models based on this mechanism. However, for different agents in temporal networks, temporal effects should be heterogeneous. Indeed, the interaction effects (such as cooperation effects in cooperation networks) between agents at different times for link-formation are different over time. 

Due to the prevalent aging effect in the network evolution process~\cite{liao_ranking_2017,dhote2013survey}, the temporal effect is typically characterized by the time-decay parameter~\cite{hu_aging_2021,kamalika_modelling_2006,zhu_effect_2003}. However, Duan et al.\ found that the contribution of behavioral synchrony indicators does not simply decay~\cite{duan2023temporal}, which may indicate that, in addition to the aging effect, other temporal effects exist in the network evolution process. So, effectively identifying the interaction effects that are strengthened or weakened at specific temporal intervals is a challenging task in understanding the network's link-formation mechanisms. The cognition-driven social network (CogSNet) model, which is proposed to simulate the changing cognition of human memory on events in social networks, involves the decay or disappearance of edge weights over time, reflecting the diminishing impact of past interactions, while they can also be strengthened through frequent interactions~\cite{michalski2021social}. Therefore, we chose the CogSNet model to characterize the heterogeneity of temporal effects in behavioral synchrony.

In social networks, human’s perception of events evolves continuously over time. The effect of behavioral synchrony changes with human cognition. Therefore, we introduce an extension of the behavioral synchrony-based link prediction model by incorporating a cognitive mechanism. This mechanism considers the evolving perception of link significance in a social network, accounting for changes in both time and intensity of interactions. Integrating a dynamic cognitive mechanism of edge importance into the behavioral synchrony-based model would enhance its ability to capture the evolving nature of relationships between nodes.  We selected several real social datasets to test the accuracy of our algorithm and compared the results with local structure-based link prediction models and competing typical temporal link prediction models.

The contributions of this paper are as follows:
\begin{enumerate}
    \item  Proposing link prediction models that incorporate social cognition mechanism and behavioral synchrony mechanism.
\item Evaluating the model's prediction ability in three scenarios: social ties, missing events, and potential links.
\item Proving the superiority of our link prediction model.
\item Distinguishing the contribution of structural similarity and temporal similarity.
\end{enumerate}

The rest of this paper is arranged as follows: Section \ref{sec: preliminaries} introduces the NSTV model and the CogSNet mechanism which will be used in our prediction models; Section \ref{sec: method} presents the six kinds of temporal link prediction model by combining cognitive neighborhood similarity and cognitive behavioral synchrony; Section \ref{sec: experiments} designs the experiments, presents datasets, and introduces evaluation metrics; Section \ref{sec: results} analyzes the results; Section \ref{sec: conclusion} discusses these prediction models.

\section{Preliminaries}
\label{sec: preliminaries}

\subsection{NSTV link prediction model}
NSTV, neighborhood-based similarity time vector link prediction model, is a kind of temporal link prediction model, which captures the behavioral synchrony among different agents over timescale and considers this synchrony as a crucial mechanism for facilitating the network link-formation~\cite{duan2023temporal}. Previous research has confirmed the crucial importance of behavioral synchrony and its underlying neural mechanisms in the formation of interpersonal relationships in social networks~\cite{pfaus_distal_2023}. Specifically, behavioral synchrony represents individuals' synchronous responses or actions within the same temporal domain in a social system, which may stem from the same external stimuli, thus reflecting a similarity among agents.

In the NSTV model, the behavior of agents within a given temporal network $G(V,E,T)$ is described as an n-dimensional temporal vector. The behavioral synchrony between agents is defined as the cosine similarity of their temporal vectors, which is called timescale similarity:
\begin{equation}
    t_{xy}=\text{Cos}(T_x,T_y)=\frac{ {\textstyle \sum_{i=1}^{k}(t_{xi}\times t_{yi})} }{\sqrt{ {\textstyle \sum_{i=1}^{k}(t_{xi})^2 } } \times \sqrt{ {\textstyle \sum_{i=1}^{k}(t_{yi})^2}}} 
    \label{nstv-cosine similarity}
\end{equation}
where $T_x = (d_{x1},d_{x2},\dots,d_{xk})$ is the temporal activity of node $v_x$ and $d_{xk}$ is the degree of node $v_x$ at time $\tau_k$.

The NSTV model added the timescale similarity to the neighborhood-based similarity link prediction models. So, the link prediction score of this model was defined as:
\begin{equation}
    \text{Sim}_{xy}^{NSTV}=\alpha \times \frac{t_{xy}}{t_\text{max}} + (1-\alpha) \times \frac{s_{xy}}{s_\text{max}}, \alpha \in [0,1]  
    \label{nstv-score}
\end{equation}
where $t_{\max}$ represents the maximum value of the timescale similarity among all node pairs, $s_{xy}$ is the score of one of the neighborhood-based link prediction models (like common neighbor), and correspondingly, $s_{\max}$ is the maximum value among all node pairs under this prediction model. Equation \eqref{nstv-score} shows the NSTV model contains two components: timescale similarity and neighborhood-based similarity, representing two distinct link-formation mechanisms. We can investigate the contributions of these two mechanisms to model prediction or other tasks by adjusting the $\alpha$ parameter.

\subsection{CogSNet -- Cognitive Social Networks}
CogSNet~\cite{michalski2021social} is a temporal network model incorporating cognitive aspects of social perception. The model explicitly portrays each social interaction as a trace in human memory, complete with its corresponding dynamics. In other words, the weights of edges in the network decay over time and may eventually vanish or be strengthened through frequent interactions -- similar to how memory imprints work in our brains. CogSNet was developed to capture interaction dynamics in social networks accurately; nevertheless, its definition is broad enough to be applied to any temporal network.  

The key elements of the model include the forgetting function and three parameters that determine the dynamics of edge weights: $\mu$ (reinforcement peak), $\theta$ (cut-off threshold), and $\lambda$ (forgetting intensity). The forgetting function over time interval $\Delta t$ can be any monotonic non-increasing function with $f(0)$ equal to $1$ and $f(t) \geq 0$ for all $t > 0$. In our study, we focused on three functions: exponential (Equation~\eqref{eqn:f_exp}), power (Equation~\eqref{eqn:f_pow}), and linear (Equation~\eqref{eqn:f_lin}). The reinforcement peak $0 < \mu \leq 1$ specifies the extent to which an event between nodes $i$ and $j$ leads to an increase in the edge weight $w_{ij}$. If the edge weight falls below the cut-off threshold (denoted as $0 < \theta < \mu$), the edge is removed.

\begin{equation}
    f^\text{exp}(\Delta t) = e^{-\lambda \Delta t}
    \label{eqn:f_exp}
\end{equation}

\begin{equation}
    f^\text{pow}(\Delta t) = \max(1,\Delta t)^{-\lambda}
    \label{eqn:f_pow}
\end{equation}

\begin{equation}
    f^\text{lin}(\Delta t) = -\Delta t\lambda
    \label{eqn:f_lin}
\end{equation}

All model parameters ($\mu, \theta, \lambda$) can be collectively summarized and presented as the lifetime $L$ of an edge for each forgetting function:

\begin{equation}
    L^\text{exp} = \frac{1}{\lambda}\ln \left( \frac{\mu}{\theta} \right)
    \label{eqn:l_exp}
\end{equation}

\begin{equation}
    L^\text{pow} = \left( \frac{\mu}{\theta} \right) ^{1/\lambda}
    \label{eqn:l_pow}
\end{equation}

\begin{equation}
    L^\text{lin} = \frac{1}{\lambda} \left( \mu - \theta \right)
    \label{eqn:l_lin}
\end{equation}

\begin{figure}
    \begin{center}
        \includegraphics[scale=0.3]{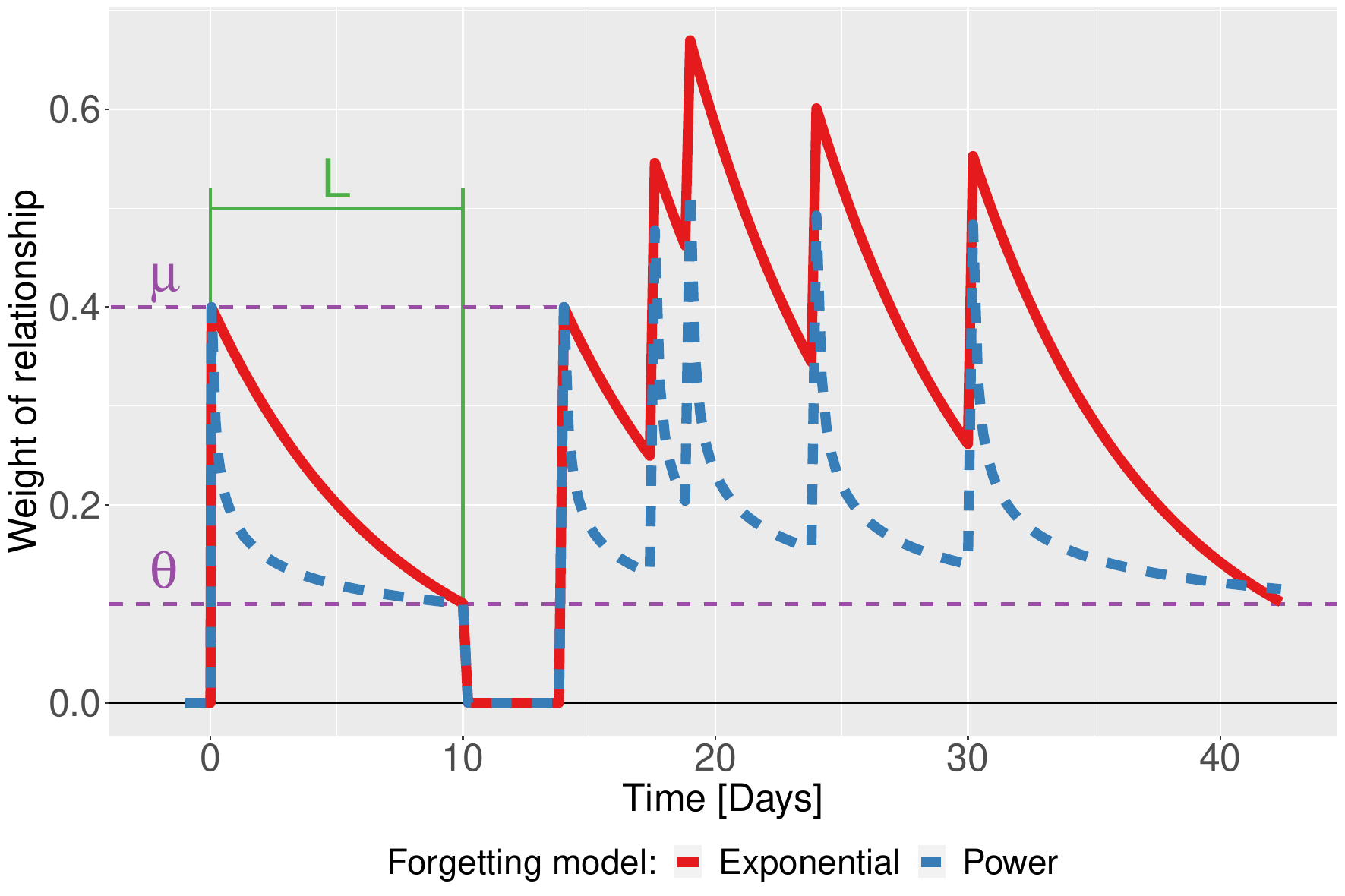}
        \caption{An example of weight between two nodes throughout the time in the CogSNet network with exponential and power functions. Parameters' values set to $\mu = 0.4$, $\theta = 0.1$, and $L = 10$ days~\cite{michalski2021social}.}
        \label{fig:cogsnet}
    \end{center}
\end{figure}

The update of the model weight when a new event occurs is defined by the following equation:
\begin{equation}
    w_{xy}(t)=\begin{cases}
        \mu,  ~~~~~~~~~~~~~~~~~~~~~~~~\text{if $w_{xy}(t_{xy})f(t-t_{xy})<\theta$},\\
        \mu + w_{xy}(t_{xy})f(t-t_{xy})(1-\mu),~~~~~~~~~ \text{otherwise},
    \end{cases}
    \label{eqn:cogsnet}
\end{equation}
    where $t_{xy}$ represents the time of the previous event for the edge $xy$ and $t$ is the time of the new event. Equation~\eqref{eqn:cogsnet} and constraints on the parameters ensure that the edge weights are within the range $[\theta,1) \cup {0}$. Additionally, the weight $w_{xy}$ can be computed for any time $t$ in the following way: $w_{xy}(t) = \nolinebreak w_{xy}(t_{xy}) f(t-t_{xy})$. Linear forgetting requires a small modification for the weight computation and update (Equation~\eqref{eqn:cogsnet}); the exact formula is shown in the appendix. An example of CogSNet's weight change mechanism for a single edge is presented in Figure~\ref{fig:cogsnet}.

\begin{figure*}[ht]
    \centering
    
    \subfloat[]{\includegraphics[width=0.79\linewidth]{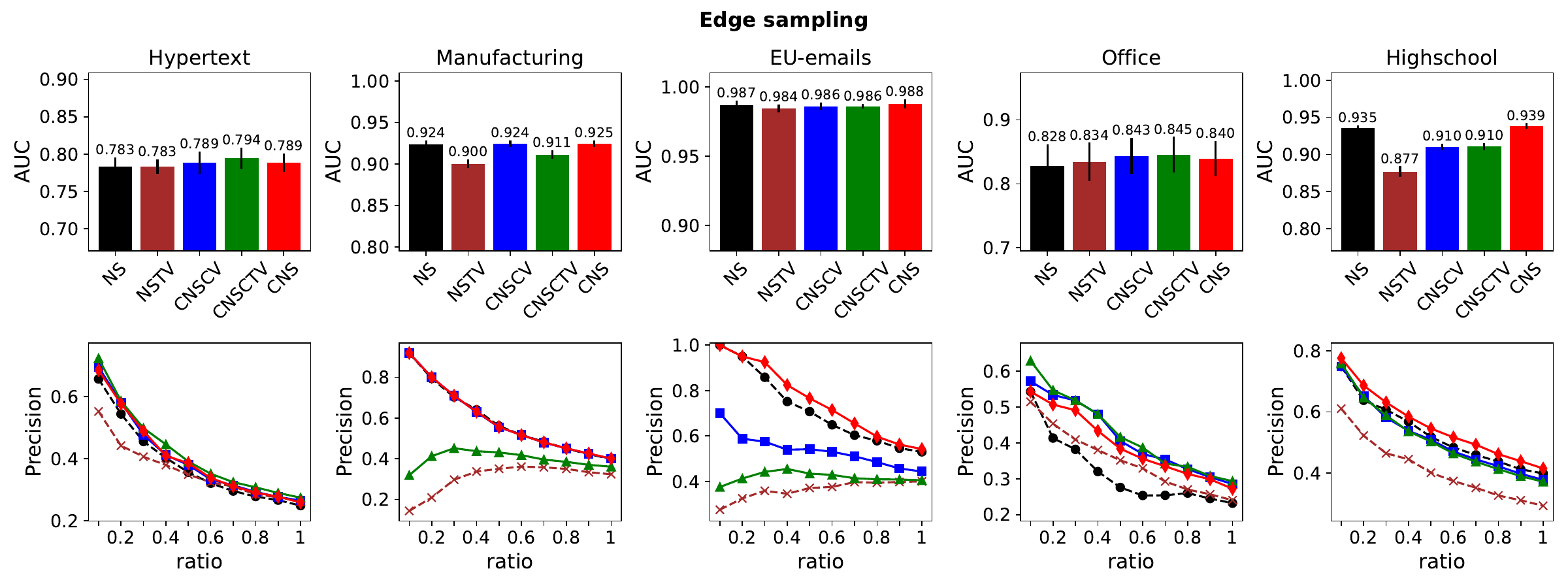}}
    
    \subfloat[]{\includegraphics[width=0.79\linewidth]{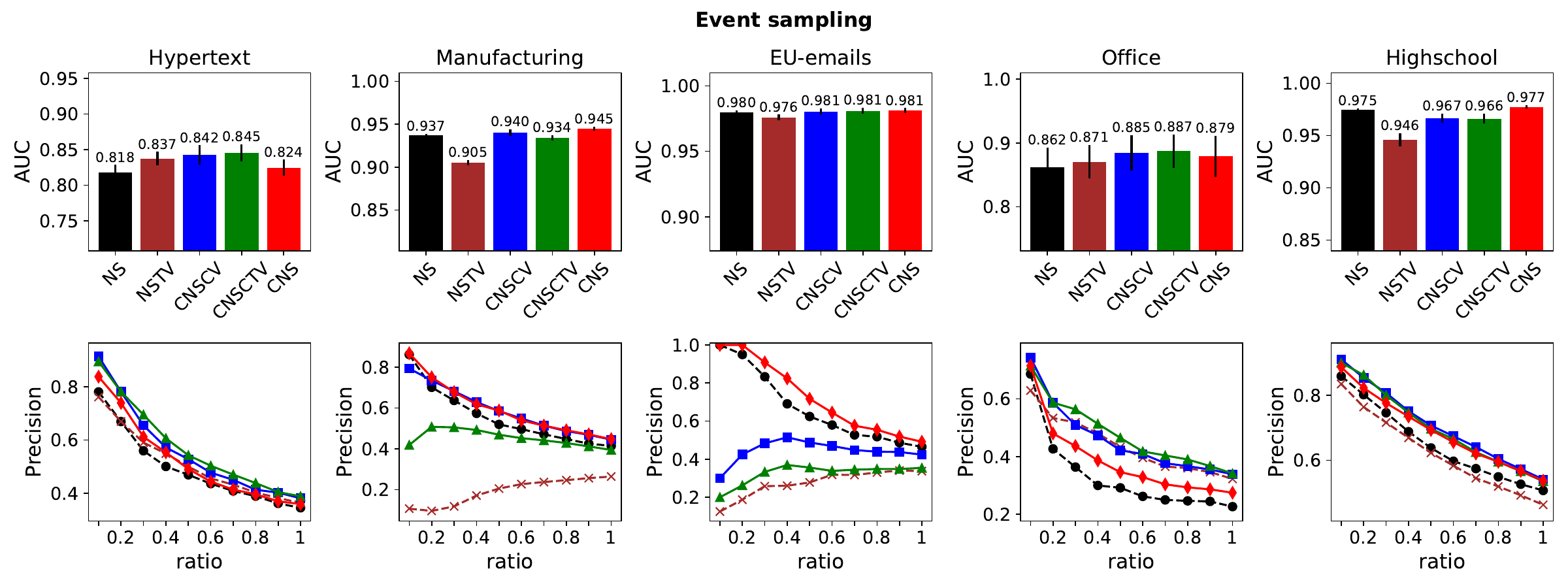}}

    \subfloat[]{\includegraphics[width=0.79\linewidth]{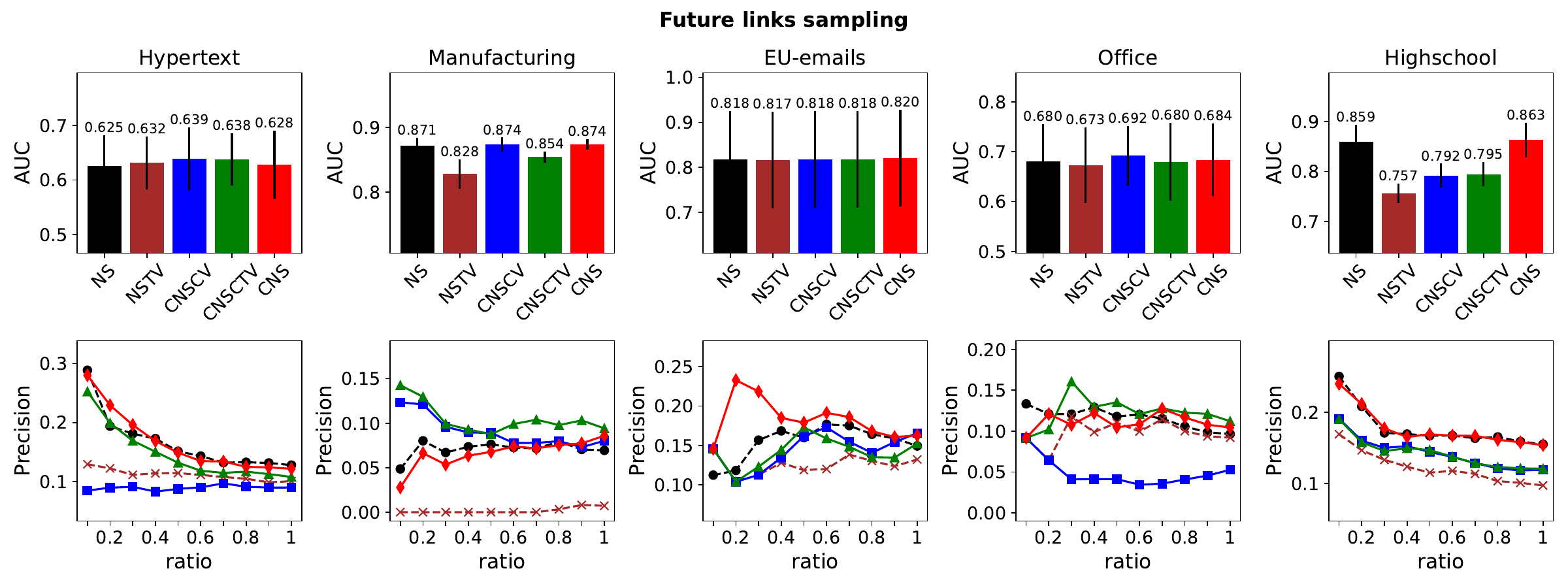}}
    
    \caption{Predictive evaluation of methods based on CogSNet when parameters were selected to maximize AUC for (a) edge sampling, (b) event sampling, and (c) future links sampling. The ratio in precision plots determines the complexity of the problem - a higher value indicates more links to predict. The precision plot uses the same colors for methods as the AUC plot.}
    \label{fig:best_auc}
\end{figure*}

\begin{figure*}[ht]
    \centering
    
    \subfloat[]{\includegraphics[width=0.79\linewidth]{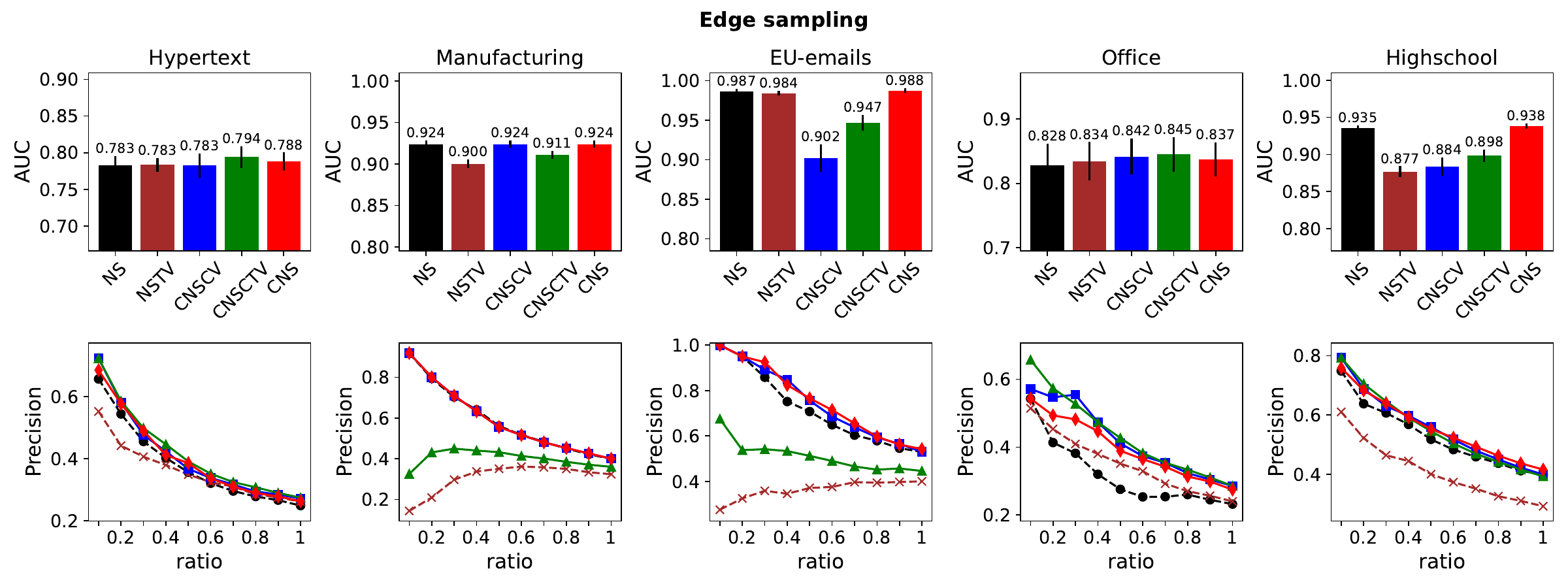}}
    
    \subfloat[]{\includegraphics[width=0.79\linewidth]{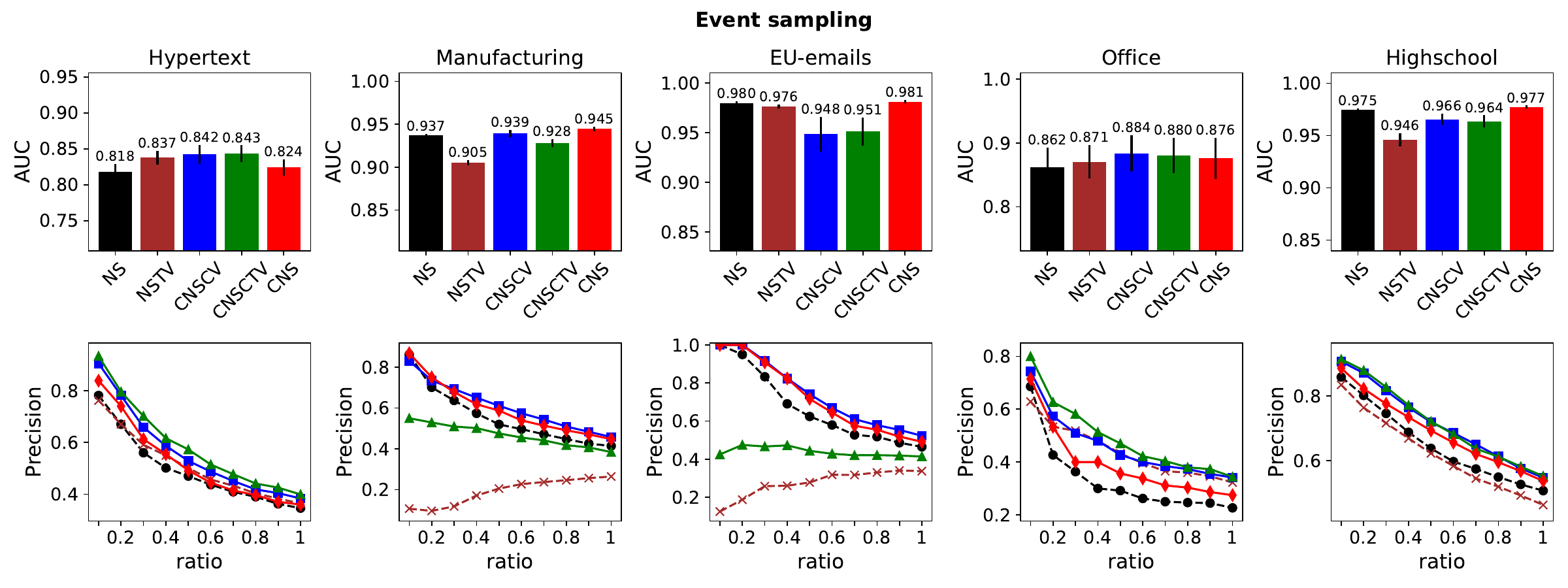}}

    \subfloat[]{\includegraphics[width=0.79\linewidth]{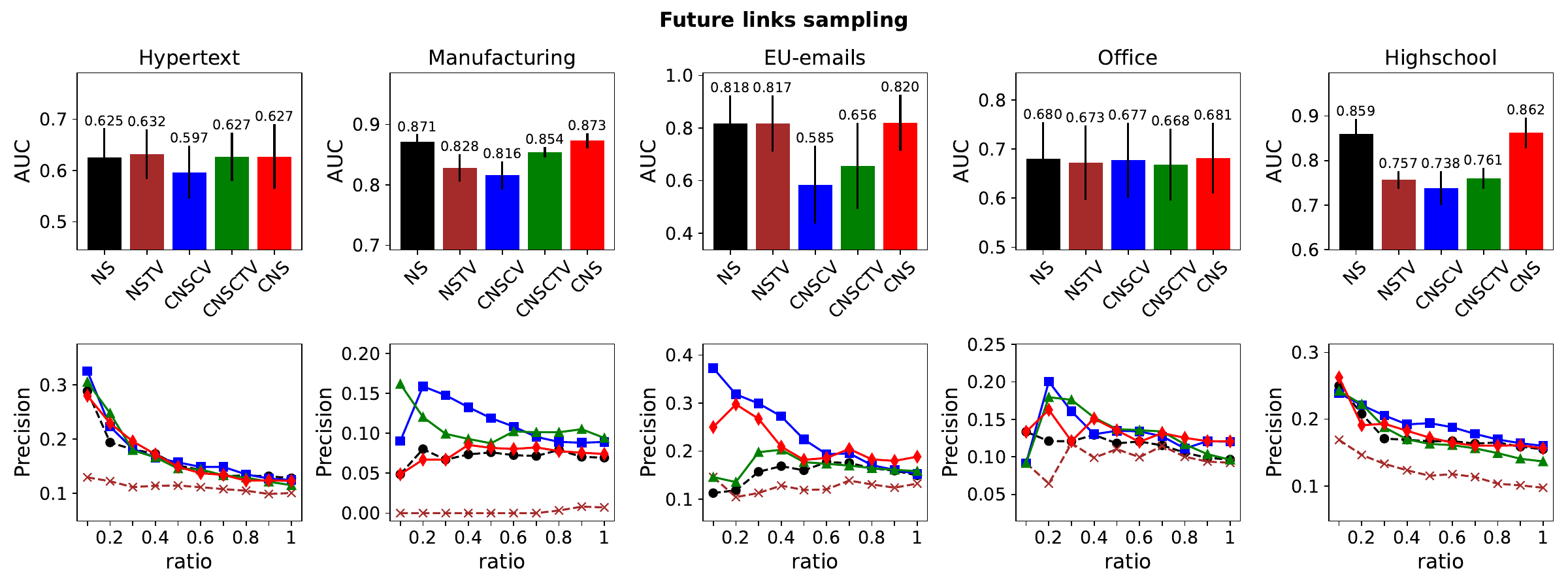}}
    
    \caption{
Predictive evaluation of methods based on CogSNet when parameters were selected to maximize precision for (a) edge sampling, (b) event sampling, and (c) future links sampling. The ratio in precision plots determines the complexity of the problem -- a higher value indicates more links to predict. The precision plot uses the same colors for methods as the AUC plot.}
    \label{fig:best_prec}
\end{figure*}

\section{Method}
\label{sec: method}
This section describes the link prediction methods proposed in this paper. First, we present a cognitive extension of neighborhood similarity and the mechanism for building cognitive vectors of node activities. Next, we propose a set of link prediction methods that combine the mechanisms of temporal behavior synchrony and the cognitive mechanism.

\subsection{Cognitive Neighbor Similarity}
The Common Neighbors (CN) measure, due to its simplicity and effectiveness in link prediction~\cite{lorrain1971structural,liben2003link}, has been used as the basis for our cognitive measure of neighborhood similarity. Calculating the number of common neighbors between two nodes is straightforward $CN_{xy} = \nolinebreak |\Gamma(x) \cup \Gamma(y)|$, where $\Gamma(x)$~is the set of neighbors of node $x$.

The CN does not take into account that some common neighbors may be more important than others -- each common neighbor is given a weight of one. Intuitively, if two unconnected nodes share a neighbor with whom an edge connects them with a high weight, they should have a higher probability than if that weight were small. In our CN extension, we used CogSNet to model a temporal network and then utilized the obtained weights to compute our Cognitive Common Neighbors (CCN) metric. The CogSNet weights take into account the frequency and time since the last interaction. CCN at time $t$ is defined as follows:

\begin{equation}
    \text{CCN}_{xy}(t) = \sum\limits_{z \in \Gamma(x, t) \cap \Gamma(y, t)} w_{xz}(t) + w_{yz}(t)
\end{equation}

\noindent where $\Gamma(x, t)$ is the set of neighbors of node $x$ at time $t$, and $w_{xz}(t)$ and $w_{yz}(t)$ are the weight from CogSNet for node pair $x,z$ and $y,z$ at time~$t$.

\subsection{Cognitive Behavioral Synchrony}

The cognitive vector $C_x$ for node $x$, describing its interaction with its neighbors, is defined as:
\begin{equation}
    C_x = [c_{x1}, c_{x2}, \dots, c_{xn}]
\end{equation}
where each element \(c_{xn}\) is computed by summation (Eq.~\eqref{eqn:cv_sum}) or averaging (Eq.~\eqref{eqn:cv_avg}) operation.
\begin{equation}
    c_{xn} = \sum\limits_{y \in \Gamma(x)} w_{xy}(t_n) A_{xy}(t_{n-1},t_{n})
    \label{eqn:cv_sum}
\end{equation}
\begin{equation}
    c_{xn} = \frac{1}{\sum\limits_{y \in \Gamma(x)}A_{xy}(t_{n-1},t_{n})}\sum\limits_{y \in \Gamma(x)} w_{xy}(t_n) A_{xy}(t_{n-1},t_{n})
    \label{eqn:cv_avg}
\end{equation}
The CogSNet weight of the edge between nodes $x$ and $y$ at time $t_n$ is represented by $w_{xy}(t_n)$, and $A_{xy}(t_{n-1}, t_n)$ is the indicator function that returns 1 if there is activity between nodes $x$ and $y$ in the time interval $[t_{n-1}, t_n]$, and 0 otherwise. The length of the vector $|C_x| = n$ depends on the size of the defined time interval. The size of the time interval between consecutive elements $c_{xn-1}$ and $c_{xn}$ is always the same. In the most extreme case, one can adopt a zero time interval, which will be equivalent to adding a new element to the cognitive vector after each event. In this scenario, the length of the vector will be equal to the number of events.

\subsection{Combining Neighborhood Similarity and Behavioral Synchrony}
Neighborhood similarity and behavioral synchrony vectors can be combined to create more complex temporal link prediction models. The previously described methods like CN, CCN, temporal vectors from NSTV, and cognitive vectors serve as building blocks that can be flexibly integrated in various combinations. In our work, we propose the following six methods:
\begin{description}
    \item[CNS]  Cognitive Neighborhood-based Similarity
    \item[NSCV]  Neighborhood-based Similarity and Cognitive Vectors
    \item[CNSCV]  Cognitive Neighborhood-based Similarity and Cognitive Vectors
    \item[NSCTV]  Neighborhood-based Similarity and Cognitive/Temporal Vector
    \item[CNSTV]  Cognitive Neighborhood-based Similarity and Temporal Vectors
    \item[CNSCTV]  Cognitive Neighborhood-based Similarity and Cognitive/Temporal Vectors
\end{description}

Here, we select CN (Common Neighbors) to represent neighborhood-based similarity (NS) and CCN (Cognitive Common Neighbors) to represent cognitive neighborhood similarity (CNS). In the NSCTV and CNSCTV methods, the behavioral synchrony vector takes the following form: $[c_{x1}, c_{x2}, \dots, c_{xn}, d_{x1}, d_{x2}, \dots d_{xk}]$, where $c_{xn}$ are elements of the cognitive vector, and $d_{xk}$ are elements of the temporal vector. Cognitive and temporal vectors capture different types of information. The former captures the dynamics of interactions (the strength of links) of a node with its neighborhood over time, while the latter contains information on how the number of neighbors with which a node communicates has changed. We summarized the individual components of the methods in Table~\ref{tab:methods}. The NS and NSTV methods in the first two lines are the baseline methods, and the remaining bold parts are prediction models proposed by us.
\begin{table}[h]
    \centering
    \caption{CogSNet methods}
    \begin{tabular}{|c|c|c|c|c|}
        \hline
         & CN & CCN & Cognitive Vector & Time Vector \\
        \hline
        NS & + & & & \\
        \hline
        NSTV & + & & & + \\
        \hline
        \textbf{CNS} & & + & & \\
        \hline
        \textbf{NSCV} & + & & + & \\
        \hline
        \textbf{CNSCV} & & + & + & \\
        \hline
        \textbf{CNSTV} & & + & & + \\
        \hline
        \textbf{NSCTV} & + & & + & + \\
        \hline
        \textbf{CNSCTV} & & + & + & + \\
        \hline
    \end{tabular}
    \label{tab:methods}
\end{table}

\section{Experiments}
\label{sec: experiments}
\subsection{Experimental Setup}

Classical link prediction can be divided into two types of problems: predicting missing links and predicting future links. In temporal social networks, each event is recorded with time information, and the interaction between two agents is not only a simple relationship of an aggregation edge~\cite{yanchenko_rev}. Thus, predicting missing links can be extended to predicting social ties and predicting missing events. These two types of problems correspond to different edge sampling methods for the test set: in the former, each edge had an equal probability of being placed in the test set -- \emph{edge sampling}. In the latter, the probability of placing an edge in the test set was directly proportional to the number of events occurring on that edge -- \emph{event sampling}. The division of the edge set into a training and test set in the case of predicting \emph{future links} is straightforward because one can split the data at one point in time~\cite{yanchenko}. Therefore, we finally tested three sampling methods: edge, event, and future link (sometimes known as \textit{ex ante} sampling), corresponding to predicting social ties, predicting missing events, and predicting potential links, respectively.

The six methods proposed by us (CNS, NSCV, CNSCV, CNSTV, NSCTV, CNSCTV) required selecting appropriate parameters for CogSNet. To select the best parameters for each dataset and sampling type, we tested a range of different parameter values and applied k-fold cross-validation. Details regarding the tested values and the best parameters are provided in Appendix~A. For methods with a parameter $\alpha$, which controls the influence of the local neighborhood component and the synchronous behavior component, we chose the value $\alpha = 0.5$, just like in the original NSTV paper~\cite{duan2023temporal}.

Firstly, the effectiveness of the proposed set of methods was compared among each other across datasets in order to obtain an overall idea about their performance in the problem of link prediction. Next, we identified the best methods and conducted individualized analyses for each problem and dataset, allowing for a more in-depth assessment of their performance. Additionally, we examined the influence of the $\alpha$ parameter on the obtained results. For the best methods and parameters from the previous step, we tested $\alpha$ parameter values from the interval $[0,1]$ with a step of 0.1.

\subsection{Datasets Description}
Five different datasets with temporal information were selected to validate the accuracy of our proposed models, labeled as (a) Hypertext, (b) Manufacturing, (c) EU\_email, (d) Office, and (e) Highshcool, respectively. 

\begin{description}
\item[Hypertext] This dataset~\cite{Isella:2011qo} was collected during the ACM Hypertext 2009 conference, where interactions among attendees during the conference were recorded using wearable radio devices, capturing the dynamic network of face-to-face contacts.
\item[Manufacturing] This dataset~\cite{nurek2020combining} includes nine months of email communication among employees of a manufacturing company located in Poland. The dataset records information such as the sender, recipient, and timestamp of email exchanges.
\item[EU\_email] This dataset~\cite{yin_local_2017, leskovec_graph_2007} originates from the email communications of a large research institution in Europe and includes anonymous information on all emails sent and received among the institution's members.
\item[Office] This dataset~\cite{NWS:9950811} was also collected face-to-face contact information among employees in a large office building in France for two consecutive weeks using wearable devices. The participation of employees was voluntary, and the overall coverage across all departments reached 69\%.
\end{description}

Similarly, \textbf{Highschool} dataset~\cite{mastrandrea_contact_2015} utilized wearable devices to gather information on face-to-face interactions among students from 9 classes in a high school in Marseille, France, over five days.

Table~\ref{tab:datasets} records the properties of these five datasets, including the number of nodes, the number of edges, the number of events, and the number of timestamps, showing the differences in the network size and time of these five datasets. Among them, \textbf{Hypertext}, \textbf{Office}, and \textbf{Highschool} datasets are all face-to-face communication networks. \textbf{Manufacturing} and \textbf{EU\_email} datasets are online communiction networks.
\begin{table}[h]
    \centering
    \caption{Datasets Desciption}
    \begin{tabular}{|c|c|c|c|c|}
        \hline
        DATASET & NODE & EDGE & EVENT & TIMESTAMP \\
        \hline
        Hypertext & 113 & 2196 & 20818 & 5246 \\
        \hline
        Manufacturing & 167 & 3250 & 82563 & 57791 \\
        \hline
        EU\_email & 142 & 833 & 47525 & 26496 \\
        \hline
        Office & 92 & 755 & 9827 & 7104 \\
        \hline
        Highschool & 327 & 5818 & 188508 & 7375 \\
        \hline
    \end{tabular}
    \label{tab:datasets}
\end{table}

\subsection{Performance Metrics}
Two performance evaluation metrics, AUC and Precision, are used to quantify the accuracy of our prediction models. We divided the datasets into training sets $E^T$ and test sets $E^P$. The link prediction algorithm will give a fraction for each unobserved link (i.e., $U-E^T$) as potential linking probabilities and form an ordered list. The AUC evaluates the algorithm's performance for the entire sorted list, while Precision focuses only on the top $L$ links.

We tested the link prediction in three scenarios by using different test set sampling methods:
\begin{enumerate}
\item Predicting social ties: We randomly selected events from the networks and removed all events between the two agents involved in these events until 10\% of the events were removed. The removed events were defined as the test set, and the remaining 90\% of the events constituted the training set.
\item Predicting missing events: We randomly selected 10\% of the events from the network and removed them. The removed events were defined as the test set, and the remaining 90\% of the events constituted the training set. 
\item Predicting future links: We used k-fold cross-validation, arranging the dataset in chronological order of events and dividing it into five groups $\left \{d_1, d_2, d_3, d_4, d_5 \right \}$. Starting from $d_2$, each group was sequentially used as the test set: training set $d_1$ -- test set $d_2$, training set $d_1+d_2$ -- test set $d_3$, training set $d_1+d_2+d_3$ -- test set $d_4$, training set $d_1+d_2+d_3+d_4$ -- test set $d_5$.
\end{enumerate}
The results of the four experiments were averaged. The division of the training and test sets for link prediction tasks (1) and (2) were also randomly repeated five times, and the results were averaged.

AUC is defined as the probability that the randomly selected missing link is given a higher score than a randomly chosen nonexistent link~\cite{lu_link_2011}. After independent comparison for n times, the AUC value can be defined as:
\begin{equation} \label{eq:AUC}
    \text{AUC} =\frac{n^\prime+0.5n^{\prime \prime}}{n}
\end{equation}
where $n^\prime$  indicates the number of times that the score of the link in the test set $E^P$  is greater than that of the edge in the non-exist set $E^N$. $n^{\prime \prime}$  indicates the number of times that the score of the link in $E^P$  is equal to that of the link in $E^N$. According to Eq.~\ref{eq:AUC}, a higher value of AUC indicates the degree to which the potential linking mechanism of the algorithm is more accurate than the random selection. $AUC=0.5$  indicates the score is random.

In the link prediction model, Precision is defined as the ratio of selected links in the test set to selected links (selecting the top-L links in an ordered list)~\cite{lu_link_2011, divakaran_temporal_2020}. Indeed, we take the top-L links as prediction links, where the $L_r$ links are right (i.e., these $L_r$ links are all from test set $E^P$), then Precision can be defined as:
\begin{equation}\label{eq:precision}
    \text{Precision} = L_r/L
\end{equation}
Therefore, higher Precision means higher prediction accuracy. We control the task difficulty of link prediction by changing the value of $L$. When $L=|E^P|$, this corresponds to the most complex scenario when we want to predict exactly the same number of links as the size of the test set. In our experiments, we calculated precision for different values of $L$ representing different ratios of the number of links in the test set. We set $L = |E^P| \times r$, where the $r$ ranged from 0.1 to 1 with a step equal to 0.1. We also computed the average precision by averaging the results for all ratios.

\section{Results}
\label{sec: results}
\subsection{General performance}
To gain a comprehensive understanding of the effectiveness of the proposed set of CogSNet-based methods, we conducted a comparative analysis of their results. To achieve this, we constructed rankings of methods (CNS, NSCV, CNSCV, CNSTV, NSCTV, CNSCTV) for each dataset, sampling type, and evaluation measure. Next, we aggregated the rankings across datasets, obtaining the average rank for each method and providing a general understanding of method effectiveness regardless of the dataset~(Table~\ref{tab:rank}).

The first observation is that there is no single method that outperforms others and obtains good results for every type of sampling and evaluation metric. However, CNSCV, CNSCTV, and CNS methods achieved significantly higher ranks, occupying usually the top half of the rankings (places from first to third). The remaining methods consistently occupied lower ranks. This outcome was consistent for each sampling type and evaluation measure used for ranking.
 
CNS performs the best among all methods for edge sampling regardless of the evaluation metric. However, its precision decreases for future links, and for event sampling, this method only achieves average results for both metrics. CNSCTV is the most universal method, being the only one that maintains an average ranking no worse than 3.0 for every type of sampling and metric. CNSCV achieves a low average ranking of 4.0 for the AUC metric for future link sampling, while in other cases, the average ranking ranges between 1.2 and 2.8.
 
To maintain the clarity of the article, further analysis will focus only on the top three methods: CNSCV, CNSCTV, and CNS. Results for the remaining methods have been included in the Appendix~B.

\begin{table}[htbp]
\centering
\caption{Average rank (lower is better) of each CogSNet-based method across all datasets for each evaluation measure and each type of sampling.}
\begin{tabular}{|c|*{6}{c|}}
\cline{2-7}
\multicolumn{1}{c|}{} & \multicolumn{6}{c|}{Average Rank} \\
\hline
\multirow{2}{*}{Methods} & \multicolumn{2}{c|}{Edge samp.} & \multicolumn{2}{c|}{Event samp.} & \multicolumn{2}{c|}{Future links samp.} \\
\cline{2-7}
& AUC & Precision & AUC & Precision & AUC & Precision \\
\hline
CNS & \textbf{2.2} & \textbf{2.3} & 3.6 & 4.2 & \textbf{2.2} & 3.6 \\
\hline
NSCV & 4.6 & 4.6 & 4.2 & 3.4 & 4.4 & 2.8 \\
\hline
CNSCV & 2.8 & \textbf{2.3} & \textbf{2.0} & 2.4 & 4.0 & \textbf{1.2} \\
\hline
CNSTV & 3.6 & 4.4 & 3.6 & 4.4 & 3.8 & 5.4 \\
\hline
NSCTV & 4.8 & 4.8 & 4.8 & 4.4 & 4.2 & 5.0 \\
\hline
CNSCTV & 3.0 & 2.6 & 2.8 & \textbf{2.2} & 2.4 & 3.0 \\
\hline
\end{tabular}
\label{tab:rank}
\end{table}

\subsection{Anaysis of the best CogSNet-based methods}
Fig.~\ref{fig:best_auc} illustrates a comparison of CogSnet-based methods, where parameters were selected to maximize AUC. For each sampling type, our model achieved a similar AUC compared to baselines. Our methods outperform the baselines in terms of precision, resulting in fewer false positives in predicted edges. Fig.~\ref{fig:best_prec} presents results for parameters chosen for the best precision. It can be observed that improving precision comes at the cost of lower AUC, particularly noticeable for CNSCV and CNSCTV methods. The selection of optimal parameters for CogSNet-based methods represents a trade-off between AUC and precision, leading to different parameter values depending on which metric we aim to maximize.

Regardless of which metric we choose for parameter selection, we can observe some common characteristics of the methods. The gradation in the difficulty of link prediction problems is observable. The best results were achieved for event sampling, followed by slightly inferior results for edge sampling. Predicting future edges was the most challenging problem of link prediction. In this setting, a significant decrease in both AUC and precision is visible. The effectiveness of methods for edge and event sampling maintains a similar trend. In other words, if a method performs well for a given dataset with edge sampling, it will also perform well for event sampling, and likewise, bad methods for edge sampling remain bad for event sampling. Regarding the prediction of future potential links, there is substantial variability in precision. For both edge and event sampling, the precision decreases with an increase in the ratio (number of predicted edges). However, in the case of future edges, there is no consistent trend, and an increase in the ratio may sometimes lead to an increase in precision. This is because of the lower AUC, where true edges may not have the highest scores and may rank lower in the list of potential edges. An increase in the ratio results in a higher number of predicted edges, thus considering these true edges at lower positions.

We noticed similar behavioral patterns in the AUC and precision metrics between the CNS and NS methods. For cases where NS outperformed other methods in AUC for datasets like Manufacturing, EU-emails, and Highschool, the results were equally good or superior for the CNS method. While CNS often surpassed NS in precision, it exhibited a similar declining trend with increasing the ratio parameter. Moreover, it is worth noting that both CNS and NS methods demonstrated robust performance for datasets with professional email communication (Manufacturing and EU-emails), regardless of whether method parameters prioritized AUC or precision. However, there was notably greater variability in performance metrics for methods based on synchronous behavior for this dataset type.

\subsection{Analyzing the $\alpha$ parameter}
The methods proposed in this paper are complementary to the neighborhood-based similarity prediction models. In other words, the link prediction model consists of local structural similarity and temporal vector similarity(including CogSnet vector and temporal vector). We selected the optimal combination of parameters for each method under each dataset and explored the two-part contribution to prediction by adjusting $\alpha$ ($\alpha \subseteq \left \{ 0.0,0.1,0.2,\dots,0.9,1.0 \right \}$). When $\alpha$ equals zero, the prediction model selects only local structural similarity. In contrast, $\alpha$ equals 1 means the prediction model selects only the temporal vector similarity.

\begin{figure*}[htbp]
    \begin{center}
        \includegraphics[width=\linewidth,scale=1.00]{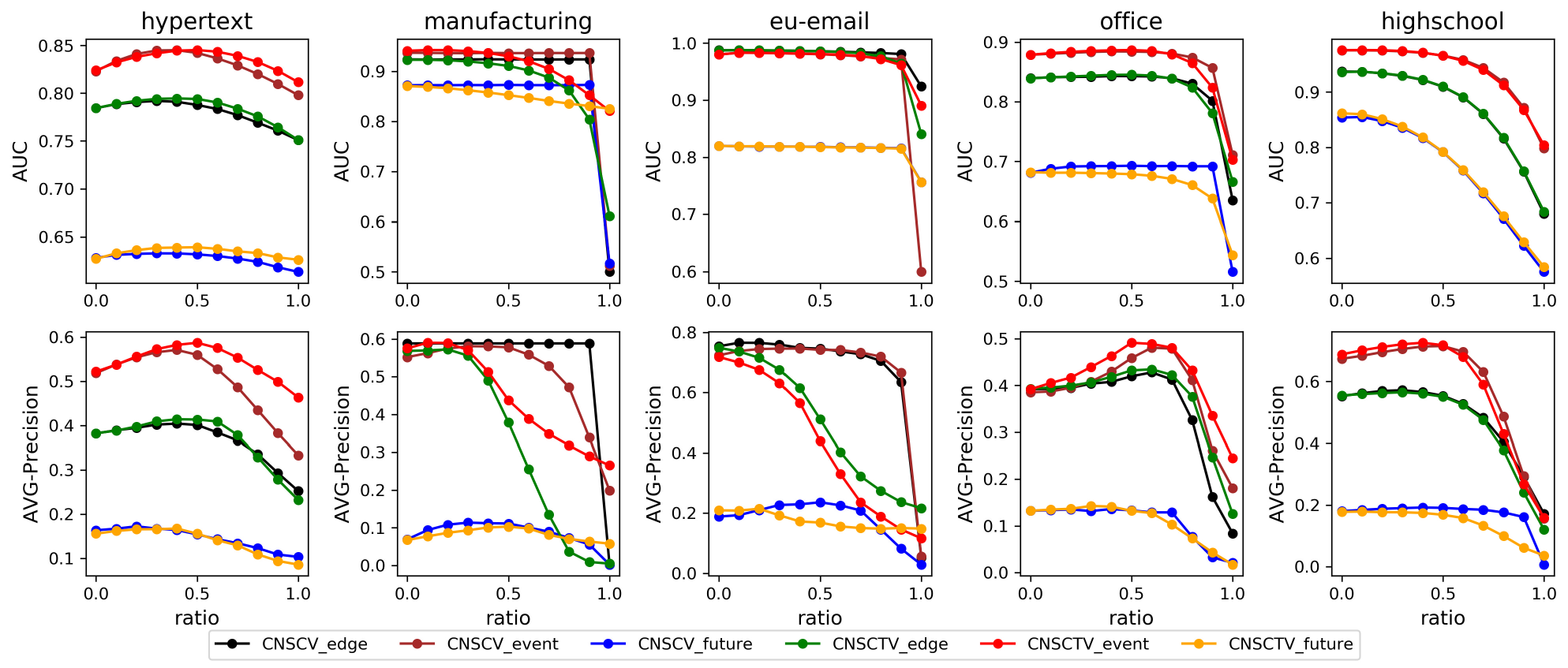}
        \caption{The contribution of timescale similarity and local structure similarity in different datasets.}
        \label{fig:alpha}
    \end{center}
\end{figure*}

Since CNS has no parameter $\alpha$, we only discuss two methods here: CNSCV and CNSCTV. Figure 4 shows that, under the measure of Precision, the different prediction models show a convex curve in most cases. The precision of the model gradually increases as the value of $\alpha$ increases, but the Precision of the model will decrease when $\alpha$ exceeds a certain value (the value is about 0.5 in most cases), which suggests that the addition of behavioral synchronization mechanism can effectively distinguish the predicted false-positive results. However, under the measure of AUC, the curve radians are not obvious. That is because, in most cases, the AUC is already close to 1 by only using neighborhood-based similarity prediction models. In other words, local structural similarity alone can identify links with greater edging potential from all unobserved links. 

We also found that the accuracy of the model decreases rapidly as $\alpha$ approaches 1. This shows that while the time factor is a crucial element in the prediction model, we cannot predict completely in isolation from the local structural similarity of the network. Previous studies have also proved that the link prediction method based on local structural similarity is effective and simple.

In the datasets, EU\_email and Manufacturing (two email networks), the improvement of model prediction accuracy is not very sensitive to the introduction of behavioral synchronization mechanisms. Different from the other three face-to-face communication networks, the link-formation mechanisms of online communication networks don’t depend on the agents' behavioral synchronization. Therefore, we can effectively select models according to the characteristics of different networks.

\subsection{CogSNet time interval analysis}
CogSNet is a continuous model that allows for creating network snapshots at any given moment, thus avoiding the aggregation of interactions and the problem of window size influence on the results~\cite{saganowski2012influence,krings2012effects}. We analyzed the impact of the snapshot time interval on the results. We can observe much greater variability in the CNSCV method than CNSCTV, both for AUC and precision for each type of sampling and dataset. The greater stability of the latter method results from the presence of an additional time vector, not just a cognitive vector. In terms of precision, in most cases, the best result is obtained for a non-zero time interval. This means that creating snapshots after each event may introduce some noise that may negatively affect link prediction. A significant gain from increasing the time interval was observed for the Manufacturing and EU-email datasets, especially for the CNSCV method. These datasets contain data from at least several months, which may capture specific patterns of edge weight behavior. In contrast, for the remaining datasets, a period of several days might be too short. Additionally, it is important to consider the different nature of datasets, where email communication may require a longer time interval than face-to-face interaction.

\begin{figure*}[htbp]
    \begin{center}
        \includegraphics[width=\linewidth,scale=1.00]{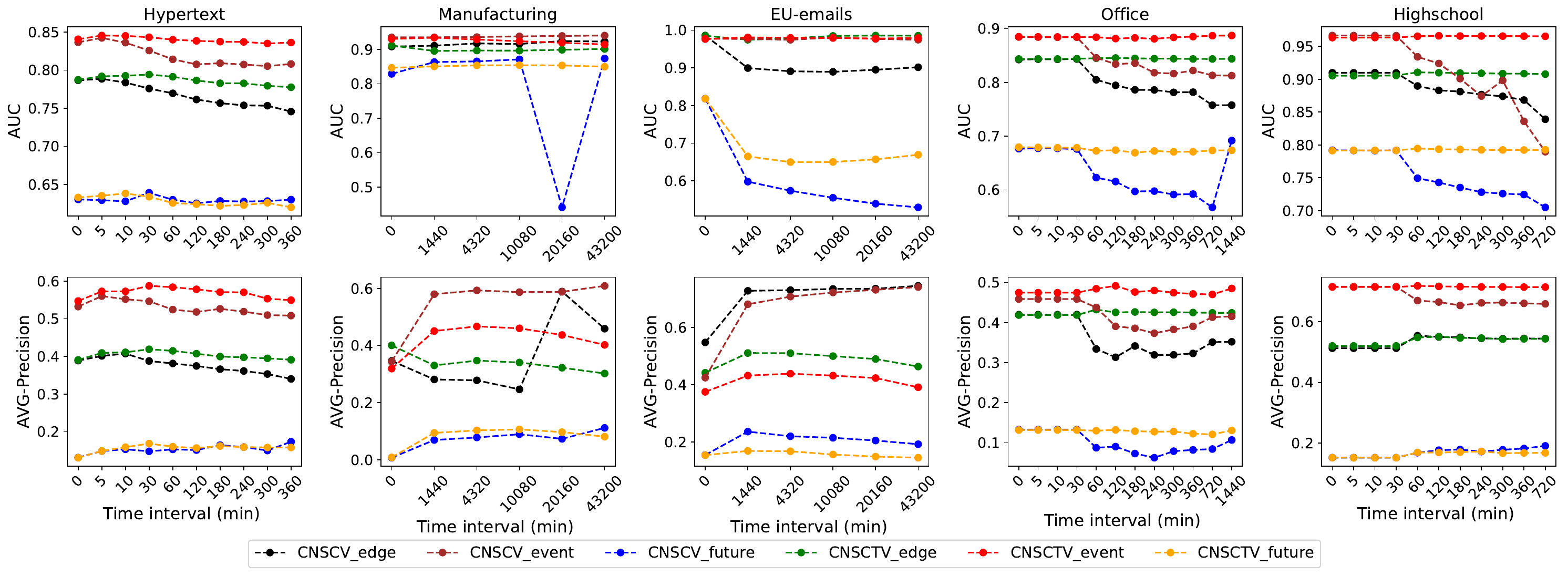}
        \caption{The impact of CogSNet snapshot time interval on results. The parameters yielding the best AUC were selected for the top row of plots, while for the bottom row, the parameters yielding the best precision were chosen. The selected parameters remained constant, with only the time interval changing.}
        \label{fig:alpha}
    \end{center}
\end{figure*}

\section{Conclusion}
\label{sec: conclusion}
This paper is an extension of the NSTV temporal link prediction model, which also discussed the importance of different temporal layers in previous work~\cite{duan2023temporal}. The authors attempted to distinguish their importance by the number of interactions at different temporal intervals. However, this improvement would fix little. The proposed temporal link prediction model in this paper more effectively identifies the heterogeneity of temporal effects by combining the CogSNet mechanism and behavioral synchrony indicators. This method not only distinguishes the importance of different temporal layers but also reflects this importance differently for each agent.

Distinguished from traditional network modeling approaches, the CogSNet model computes network weights by simulating human memory's perception of events~\cite{michalski2021social}, making it more suitable for social network modeling. This paper proposes a series of link prediction methods based on the CogSNet mechanism and agents’ behavioral synchronization. We attempt to combine the mechanisms of neighborhood-based similarity, CogSNet perception, and behavioral synchronization. The results show that using CogSNet-based network weights for calculating neighborhood-based similarity and behavioral synchronization in link prediction models improves performance across different scenarios and evaluation metrics. Previous studies have indicated that solely relying on time decay effects for network weight resetting does not significantly enhance the accuracy of temporal link prediction models~\cite{duan2023temporal}. Our models demonstrate that the CogSNet model better simulates the perception of events by human agents in social networks.

The innovation of this paper lies in: firstly, we compute network weights through the CogSNet model to better capture social network characteristics. Secondly, under the weights reset network, we propose a series of temporal link prediction models based on local structure and behavioral synchronization, outperforming existing models in terms of AUC and Precision. Moreover, we attempt to focus on different scenarios of link prediction, including predicting social ties, missing events, and potential links. Among them, the models for predicting social ties and missing events have matured with high AUC and Precision in prediction models. However, predicting future potential links remains a challenging problem\cite{yanchenko_rev}.

In the prediction of different datasets, we find that both agents’ behavioral synchronization and the CogSNet event perception mechanism have strong effects in face-to-face communication networks, but mediocre in online email networks. Especially in the allocation of contributions to local structure and temporal vectors, the prediction model relies more on structure. That is to say, time effects in networks cannot serve as sole predictors but can extend and complement prediction indicators based on static local structures. We observed similar patterns in the same type of datasets. Therefore, determining how to set parameter combinations in prediction models for datasets with similar characteristics will be the focus of our future research.

\section{Acknowledgment}
Y.D. and Q.G. were supported by funding from the National Natural Science Foundation of China (Grant No.\ 42001236), Young Elite Scientist Sponsorship Program by Bast (Grant No.\ BYESS2023413), and the National Natural Science Foundation of China (Grant No. 71991481, 71991480). Y.D was supported by the China Scholarship Council (CSC) program (No.\ 202106400071). M.N. and R.M. were also supported in part by the National Science Centre, Poland, under Grant 2021/41/B/HS6/02798. This work was also partially funded by the Polish Ministry of Education and Science within the programme “International Projects Co-Funded”  and the European Union under the Horizon Europe, grant no. 101086321 (OMINO) -- M.N. and R.M. Views and opinions expressed are however those of the author(s) only and do not necessarily reflect those of the European Union or the European Research Executive Agency. Neither the European Union nor European Research Executive Agency can be held responsible for them. P.H. was supported by JSPS KAKENHI Grant Number JP 21H04595.

\section{Author contribution}
Y. Duan and M. Nurek contributed equally to the concept, design, methodology, dataset preprocessing, code, results analysis, and original draft writing. Q. Guan, R. Michalski, and P. Holme contributed to writing, reviewing \& editing.

\section{Data availability statement}
The code is available online at \url{https://github.com/nurek-mateusz/cognitive-temporal-link-prediction}. The used datasets are publicly available online, and references to them are provided in \textit{IV.B Data Description} section.

\bibliographystyle{abbrv}
\bibliography{references}

\appendix

\section{Modification for linear forgetting}
The linear forgetting function requires modification of the formulas for weight computation and weight update, changing from multiplying the old weight by the forgetting function value to adding the forgetting function value. The linear forgetting function is defined as . The weight update equation is defined as following:
\begin{equation}
    w_{xy}(t)=\begin{cases}
        \mu,  ~~~~~~~~~~~~~~~~~~~~~~~~~~~\text{if $w_{xy}(t_{xy})+f^\text{lin}(t-t_{xy})<\theta$},\\
        \mu + (w_{xy}(t_{xy})+f^\text{lin}(t-t_{xy}))(1-\mu),~~~~~~~~~ \text{otherwise},
    \end{cases}
    \label{eqn:appendix_cogsnet}
\end{equation}
where $t_{xy}$ represents the time of the previous event for the edge $xy$, $t$ is the time of the new event, $f^\text{lin}(t-t_{xy}) = -\Delta t \lambda$ and $f^\text{lin}(t-t_{xy}) \leq 0$. The weight $w_{xy}$ can be computed for any time $t$ in the following way: $w_{xy}(t) = \nolinebreak w_{xy}(t_{xy})+f^\text{lin}(t-t_{xy})$.

\section{CogSNet parameters}
Tab~\ref{tab:appendix_params} contains the range of tested CogSNet parameters for all methods and datasets. A value of 0 interval indicates adding an element to the cognitive vector after each event, just as it was done in the original NSTV method.

Tab~\ref{tab:appendix_best_param_auc_edge}-\ref{tab:appendix_best_param_prec_future} contain the best CogSNet parameters for each method and dataset for AUC and precision metrics.

\begin{table}[htbp]
    \centering
    \caption{The range of CogSNet parameters tested for each dataset}
    \begin{tabular}{|c|c|}
        \hline
        \multicolumn{2}{|c|}{Hypertext} \\
        \hline
        Interval (minutes) & 0, 5, 10, 30, 60, 120, 180, 240, 300, 360 \\
        $L$ (hours) & 0.5, 1, 1.5, 2, 3, 6, 12, 24 48 \\
        $\mu$ & 0.3, 0.5, 0.8 \\
        $\theta$ & 0.1, 0.2 \\
        forgetting fun. & linear, exp, pow \\
        aggregation & sum, avg \\
        \hline
        \hline
        \multicolumn{2}{|c|}{Manufacturing and EU-emails} \\
        \hline
        Interval (minutes) & 0, 1440, 4320, 10080, 20160, 43200 \\
        $L$ (hours) & 24, 72, 168, 336, 720, 2160, 4320 \\
        $\mu$ & 0.3, 0.5, 0.8 \\
        $\theta$ & 0.1, 0.2 \\
        forgetting fun. & linear, exp, pow \\
        aggregation & sum, avg \\
        \hline
        \hline
        \multicolumn{2}{|c|}{Office} \\
        \hline
        Interval (minutes) & 0, 5, 10, 30, 60, 120, 180, 240, 300, 360, 720, 1440 \\
        $L$ (hours) & 1, 1.5, 2, 3, 6, 12, 24, 48, 72, 96, 120, 144, 168, 192 \\
        $\mu$ & 0.3, 0.5, 0.8 \\
        $\theta$ & 0.1, 0.2 \\
        forgetting fun. & linear, exp, pow \\
        aggregation & sum, avg \\
        \hline
        \hline
        \multicolumn{2}{|c|}{Highschool} \\
        \hline
        Interval (minutes) & 0, 5, 10, 30, 60, 120, 180, 240, 300, 360, 720 \\
        $L$ (hours) & 1, 1.5, 2, 3, 6, 12, 24, 48, 72 \\
        $\mu$ & 0.3, 0.5, 0.8 \\
        $\theta$ & 0.1, 0.2 \\
        forgetting fun. & linear, exp, pow \\
        aggregation & sum, avg \\
        \hline
    \end{tabular}
    \label{tab:appendix_params}
\end{table}

\begin{table}[htbp]
    \centering
    \caption{The best CogSNet parameters for each method and dataset for AUC metric and edge sampling.}
    \begin{tabular}{|c|c|c|c|c|c|c|}
        \hline
        \multicolumn{7}{|c|}{AUC - EDGE SAMPLING} \\
        \hline
        \hline
        \multicolumn{7}{|c|}{Hypertext} \\
        \hline
          & CNS & NSCV & CNSCV & CNSTV & NSCTV & CNSCTV \\
        \hline
        Interval & - & 5 & 5 & 120 & 30 & 30 \\
        $L$ & 2 & 12 & 48 & 0.5 & 1.5 & 48 \\
        $\mu$ & 0.5 & 0.8 & 0.8 & 0.8 & 0.8 & 0.5 \\
        $\theta$ & 0.1 & 0.2 & 0.2 & 0.1 & 0.2 & 0.2 \\
        forget. & exp & exp & lin & pow & exp & lin \\
        agg. & sum & avg & avg & sum & sum & sum \\
        \hline
        \hline
        \multicolumn{7}{|c|}{Manufacturing} \\
        \hline
          & CNS & NSCV & CNSCV & CNSTV & NSCTV & CNSCTV \\
        \hline
        Interval & - & 20160 & 20160 & 43200 & 0 & 0 \\
        $L$ & 2160 & 2160 & 2160 & 24 & 2160 & 4320 \\
        $\mu$ & 0.5 & 0.3 & 0.5 & 0.3 & 0.3 & 0.3 \\
        $\theta$ & 0.1 & 0.2 & 0.1 & 0.1 & 0.2 & 0.2 \\
        forget. & pow & pow & pow & lin & lin & lin \\
        agg. & sum & sum & avg & sum & sum & avg \\
        \hline
        \hline
        \multicolumn{7}{|c|}{EU-emails} \\
        \hline
          & CNS & NSCV & CNSCV & CNSTV & NSCTV & CNSCTV \\
        \hline
        Interval & - & 0 & 0 & 1440 & 43200 & 20160 \\
        $L$ & 4320 & 24 & 4320 & 4320 & 168 & 336 \\
        $\mu$ & 0.3 & 0.3 & 0.3 & 0.3 & 0.3 & 0.8 \\
        $\theta$ & 0.2 & 0.1 & 0.2 & 0.2 & 0.2 & 0.1 \\
        forget. & lin & lin & exp & lin & lin & exp \\
        agg. & sum & avg & avg & sum & avg & avg \\
        \hline
        \hline
        \multicolumn{7}{|c|}{Office} \\
        \hline
          & CNS & NSCV & CNSCV & CNSTV & NSCTV & CNSCTV \\
        \hline
        Interval & - & 10 & 30 & 360 & 240 & 120 \\
        $L$ & 168 & 1.5 & 168 & 192 & 2 & 168 \\
        $\mu$ & 0.3 & 0.8 & 0.3 & 0.3 & 0.8 & 0.3 \\
        $\theta$ & 0.2 & 0.1 & 0.2 & 0.2 & 0.2 & 0.2 \\
        forget. & lin & lin & lin & lin & pow & lin \\
        agg. & sum & sum & sum & sum & avg & avg \\
        \hline
        \hline
        \multicolumn{7}{|c|}{Highschool} \\
        \hline
          & CNS & NSCV & CNSCV & CNSTV & NSCTV & CNSCTV \\
        \hline
        Interval & - & 30 & 10 & 30 & 60 & 60 \\
        $L$ & 48 & 6 & 2 & 2 & 48 & 12 \\
        $\mu$ & 0.3 & 0.3 & 0.3 & 0.3 & 0.8 & 0.8 \\
        $\theta$ & 0.2 & 0.2 & 0.1 & 0.2 & 0.2 & 0.2 \\
        forget. & lin & pow & pow & exp & exp & lin \\
        agg. & sum & sum & sum & sum & avg & avg \\
        \hline
        \hline
    \end{tabular}
    \label{tab:appendix_best_param_auc_edge}
\end{table}

\begin{table}[htbp]
    \centering
    \caption{The best CogSNet parameters for each method and dataset for AUC metric and event sampling.}
    \begin{tabular}{|c|c|c|c|c|c|c|}
        \hline
        \multicolumn{7}{|c|}{AUC - EVENT SAMPLING} \\
        \hline
        \hline
        \multicolumn{7}{|c|}{Hypertext} \\
        \hline
          & CNS & NSCV & CNSCV & CNSTV & NSCTV & CNSCTV \\
        \hline
        Interval & - & 5 & 5 & 240 & 300 & 5 \\
        $L$ & 48 & 12 & 12 & 48 & 3 & 48 \\
        $\mu$ & 0.3 & 0.8 & 0.3 & 0.3 & 0.3 & 0.3 \\
        $\theta$ & 0.2 & 0.2 & 0.2 & 0.2 & 0.1 & 0.2 \\
        forget. & lin & lin & lin & lin & exp & lin \\
        agg. & sum & avg & sum & sum & sum & sum \\
        \hline
        \hline
        \multicolumn{7}{|c|}{Manufacturing} \\
        \hline
          & CNS & NSCV & CNSCV & CNSTV & NSCTV & CNSCTV \\
        \hline
        Interval & - & 20160 & 43200 & 43200 & 1440 & 1440 \\
        $L$ & 2160 & 336 & 720 & 4320 & 168 & 720 \\
        $\mu$ & 0.5 & 0.8 & 0.3 & 0.3 & 0.3 & 0.3 \\
        $\theta$ & 0.2 & 0.1 & 0.1 & 0.1 & 0.1 & 0.2 \\
        forget. & lin & lin & exp & exp & exp & lin \\
        agg. & sum & avg & avg & sum & sum & sum \\
        \hline
        \hline
        \multicolumn{7}{|c|}{EU-emails} \\
        \hline
          & CNS & NSCV & CNSCV & CNSTV & NSCTV & CNSCTV \\
        \hline
        Interval & - & 10080 & 10080 & 1440 & 43200 & 1440 \\
        $L$ & 4320 & 4320 & 336 & 4320 & 72 & 24 \\
        $\mu$ & 0.3 & 0.8 & 0.3 & 0.3 & 0.3 & 0.8 \\
        $\theta$ & 0.2 & 0.1 & 0.2 & 0.2 & 0.2 & 0.1 \\
        forget. & lin & pow & pow & lin & lin & exp \\
        agg. & sum & avg & avg & sum & avg & sum \\
        \hline
        \hline
        \multicolumn{7}{|c|}{Office} \\
        \hline
          & CNS & NSCV & CNSCV & CNSTV & NSCTV & CNSCTV \\
        \hline
        Interval & - & 0 & 0 & 5 & 720 & 1440 \\
        $L$ & 168 & 1.5 & 192 & 96 & 24 & 192 \\
        $\mu$ & 0.3 & 0.5 & 0.3 & 0.3 & 0.5 & 0.3 \\
        $\theta$ & 0.2 & 0.2 & 0.2 & 0.2 & 0.1 & 0.1 \\
        forget. & lin & pow & lin & lin & lin & lin \\
        agg. & sum & sum & sum & sum & avg & avg \\
        \hline
        \hline
        \multicolumn{7}{|c|}{Highschool} \\
        \hline
          & CNS & NSCV & CNSCV & CNSTV & NSCTV & CNSCTV \\
        \hline
        Interval & - & 30 & 0 & 300 & 120 & 120 \\
        $L$ & 72 & 1.5 & 12 & 12 & 72 & 48 \\
        $\mu$ & 0.3 & 0.3 & 0.3 & 0.3 & 0.8 & 0.8 \\
        $\theta$ & 0.1 & 0.1 & 0.2 & 0.1 & 0.1 & 0.2 \\
        forget. & lin & pow & pow & exp & lin & lin \\
        agg. & sum & avg & avg & sum & avg & avg \\
        \hline
        \hline
    \end{tabular}
    \label{tab:appendix_best_param_auc_event}
\end{table}

\begin{table}[htbp]
    \centering
    \caption{The best CogSNet parameters for each method and dataset for AUC metric and future link sampling.}
    \begin{tabular}{|c|c|c|c|c|c|c|}
        \hline
        \multicolumn{7}{|c|}{AUC - FUTURE LINK SAMPLING} \\
        \hline
        \hline
        \multicolumn{7}{|c|}{Hypertext} \\
        \hline
          & CNS & NSCV & CNSCV & CNSTV & NSCTV & CNSCTV \\
        \hline
        Interval & - & 30 & 30 & 0 & 10 & 10 \\
        $L$ & 48 & 1.5 & 1.5 & 48 & 48 & 48 \\
        $\mu$ & 0.3 & 0.3 & 0.3 & 0.3 & 0.3 & 0.3 \\
        $\theta$ & 0.1 & 0.2 & 0.2 & 0.2 & 0.1 & 0.2 \\
        forget. & lin & pow & pow & exp & lin & lin \\
        agg. & sum & sum & sum & sum & sum & sum \\
        \hline
        \hline
        \multicolumn{7}{|c|}{Manufacturing} \\
        \hline
          & CNS & NSCV & CNSCV & CNSTV & NSCTV & CNSCTV \\
        \hline
        Interval & - & 43200 & 43200 & 0 & 4320 & 10080 \\
        $L$ & 72.0 & 4320 & 24 & 336 & 4320 & 4320 \\
        $\mu$ & 0.5 & 0.5 & 0.3 & 0.3 & 0.5 & 0.3 \\
        $\theta$ & 0.2 & 0.2 & 0.2 & 0.2 & 0.1 & 0.1 \\
        forget. & pow & pow & pow & lin & lin & exp \\
        agg. & sum & sum & avg & sum & sum & sum \\
        \hline
        \hline
        \multicolumn{7}{|c|}{EU-emails} \\
        \hline
          & CNS & NSCV & CNSCV & CNSTV & NSCTV & CNSCTV \\
        \hline
        Interval & - & 0 & 0 & 20160 & 0 & 0 \\
        $L$ & 336 & 72 & 720 & 72 & 336 & 720 \\
        $\mu$ & 0.3 & 0.3 & 0.3 & 0.3 & 0.5 & 0.3 \\
        $\theta$ & 0.2 & 0.1 & 0.2 & 0.2 & 0.1 & 0.1 \\
        forget. & exp & lin & exp & exp & pow & exp \\
        agg. & sum & sum & sum & sum & sum & sum \\
        \hline
        \hline
        \multicolumn{7}{|c|}{Office} \\
        \hline
          & CNS & NSCV & CNSCV & CNSTV & NSCTV & CNSCTV \\
        \hline
        Interval & - & 720 & 1440 & 1440 & 10 & 0 \\
        $L$ & 24 & 1 & 12 & 72 & 1 & 144 \\
        $\mu$ & 0.3 & 0.5 & 0.8 & 0.3 & 0.3 & 0.3 \\
        $\theta$ & 0.1 & 0.1 & 0.2 & 0.2 & 0.2 & 0.2 \\
        forget. & pow & lin & exp & lin & lin & lin \\
        agg. & sum & sum & sum & sum & sum & avg \\
        \hline
        \hline
        \multicolumn{7}{|c|}{Highschool} \\
        \hline
          & CNS & NSCV & CNSCV & CNSTV & NSCTV & CNSCTV \\
        \hline
        Interval & - & 0 & 0 & 360 & 60 & 60 \\
        $L$ & 72 & 1.5 & 3 & 3 & 48 & 3 \\
        $\mu$ & 0.3 & 0.3 & 0.8 & 0.8 & 0.8 & 0.8 \\
        $\theta$ & 0.2 & 0.2 & 0.2 & 0.2 & 0.2 & 0.1 \\
        forget. & lin & pow & lin & lin & exp & lin \\
        agg. & sum & sum & sum & sum & avg & avg \\
        \hline
        \hline
    \end{tabular}
    \label{tab:appendix_best_param_auc_future}
\end{table}

\begin{table}[htbp]
    \centering
    \caption{The best CogSNet parameters for each method and dataset for precision metric and edge sampling.}
    \begin{tabular}{|c|c|c|c|c|c|c|}
        \hline
        \multicolumn{7}{|c|}{PRECISION - EDGE SAMPLING} \\
        \hline
        \hline
        \multicolumn{7}{|c|}{Hypertext} \\
        \hline
          & CNS & NSCV & CNSCV & CNSTV & NSCTV & CNSCTV \\
        \hline
        Interval & - & 5 & 10 & 0 & 30 & 30 \\
        $L$ & 12 & 1.5 & 24 & 12 & 1.5 & 48 \\
        $\mu$ & 0.8 & 0.8 & 0.5 & 0.8 & 0.8 & 0.5 \\
        $\theta$ & 0.1 & 0.2 & 0.1 & 0.1 & 0.2 & 0.2 \\
        forget. & pow & lin & lin & pow & exp & lin \\
        agg. & sum & avg & avg & sum & sum & sum \\
        \hline
        \hline
        \multicolumn{7}{|c|}{Manufacturing} \\
        \hline
          & CNS & NSCV & CNSCV & CNSTV & NSCTV & CNSCTV \\
        \hline
        Interval & - & 20160 & 20160 & 0 & 0 & 0 \\
        $L$ & 168 & 24 & 336 & 168 & 4320 & 4320 \\
        $\mu$ & 0.5 & 0.3 & 0.8 & 0.3 & 0.5 & 0.3 \\
        $\theta$ & 0.1 & 0.2 & 0.2 & 0.1 & 0.2 & 0.2 \\
        forget. & pow & pow & pow & lin & lin & exp \\
        agg. & sum & sum & sum & sum & sum & sum \\
        \hline
        \hline
        \multicolumn{7}{|c|}{EU-emails} \\
        \hline
          & CNS & NSCV & CNSCV & CNSTV & NSCTV & CNSCTV \\
        \hline
        Interval & - & 43200 & 43200 & 10080 & 1440 & 1440 \\
        $L$ & 4320 & 4320 & 4320 & 4320 & 336 & 2160 \\
        $\mu$ & 0.3 & 0.5 & 0.3 & 0.3 & 0.5 & 0.3 \\
        $\theta$ & 0.2 & 0.1 & 0.2 & 0.2 & 0.1 & 0.2 \\
        forget. & lin & lin & lin & lin & lin & lin \\
        agg. & sum & avg & avg & sum & sum & sum \\
        \hline
        \hline
        \multicolumn{7}{|c|}{Office} \\
        \hline
          & CNS & NSCV & CNSCV & CNSTV & NSCTV & CNSCTV \\
        \hline
        Interval & - & 5 & 5 & 5 & 240 & 60 \\
        $L$ & 48 & 6 & 96 & 120 & 3 & 168 \\
        $\mu$ & 0.3 & 0.5 & 0.3 & 0.3 & 0.8 & 0.3 \\
        $\theta$ & 0.2 & 0.1 & 0.2 & 0.1 & 0.2 & 0.2 \\
        forget. & lin & pow & lin & lin & pow & lin \\
        agg. & sum & avg & avg & sum & sum & avg \\
        \hline
        \hline
        \multicolumn{7}{|c|}{Highschool} \\
        \hline
          & CNS & NSCV & CNSCV & CNSTV & NSCTV & CNSCTV \\
        \hline
        Interval & - & 60 & 60 & 5 & 60 & 120 \\
        $L$ & 1.5 & 6 & 24 & 1.5 & 24 & 72 \\
        $\mu$ & 0.3 & 0.3 & 0.3 & 0.3 & 0.8 & 0.3 \\
        $\theta$ & 0.2 & 0.2 & 0.1 & 0.2 & 0.1 & 0.1 \\
        forget. & pow & pow & lin & pow & lin & lin \\
        agg. & sum & sum & sum & sum & sum & sum \\
        \hline
        \hline
    \end{tabular}
    \label{tab:appendix_best_param_prec_edge}
\end{table}

\begin{table}[htbp]
    \centering
    \caption{The best CogSNet parameters for each method and dataset for precision metric and event sampling.}
    \begin{tabular}{|c|c|c|c|c|c|c|}
        \hline
        \multicolumn{7}{|c|}{PRECISION - EVENT SAMPLING} \\
        \hline
        \hline
        \multicolumn{7}{|c|}{Hypertext} \\
        \hline
          & CNS & NSCV & CNSCV & CNSTV & NSCTV & CNSCTV \\
        \hline
        Interval & - & 10 & 5 & 30 & 180 & 30 \\
        $L$ & 48 & 48 & 24 & 48 & 3 & 48 \\
        $\mu$ & 0.3 & 0.5 & 0.3 & 0.3 & 0.3 & 0.3 \\
        $\theta$ & 0.2 & 0.1 & 0.1 & 0.2 & 0.1 & 0.2 \\
        forget. & lin & lin & lin & lin & lin & lin \\
        agg. & sum & avg & sum & sum & sum & sum \\
        \hline
        \hline
        \multicolumn{7}{|c|}{Manufacturing} \\
        \hline
          & CNS & NSCV & CNSCV & CNSTV & NSCTV & CNSCTV \\
        \hline
        Interval & - & 43200 & 43200 & 20160 & 4320 & 4320 \\
        $L$ & 4320 & 336 & 168 & 2160 & 4320 & 4320 \\
        $\mu$ & 0.5 & 0.8 & 0.3 & 0.3 & 0.3 & 0.3 \\
        $\theta$ & 0.2 & 0.2 & 0.2 & 0.2 & 0.1 & 0.2 \\
        forget. & pow & lin & exp & pow & lin & lin \\
        agg. & sum & avg & avg & sum & sum & sum \\
        \hline
        \hline
        \multicolumn{7}{|c|}{EU-emails} \\
        \hline
          & CNS & NSCV & CNSCV & CNSTV & NSCTV & CNSCTV \\
        \hline
        Interval & - & 43200 & 43200 & 10080 & 4320 & 4320 \\
        $L$ & 4320 & 2160 & 4320 & 2160 & 2160 & 4320 \\
        $\mu$ & 0.3 & 0.5 & 0.3 & 0.3 & 0.3 & 0.3 \\
        $\theta$ & 0.2 & 0.1 & 0.2 & 0.2 & 0.1 & 0.1 \\
        forget. & lin & lin & lin & lin & lin & lin \\
        agg. & sum & avg & avg & sum & sum & sum \\
        \hline
        \hline
        \multicolumn{7}{|c|}{Office} \\
        \hline
          & CNS & NSCV & CNSCV & CNSTV & NSCTV & CNSCTV \\
        \hline
        Interval & - & 5 & 5 & 5 & 240 & 120 \\
        $L$ & 1.5 & 12 & 144 & 1.5 & 120 & 1.5 \\
        $\mu$ & 0.3 & 0.3 & 0.3 & 0.3 & 0.5 & 0.3 \\
        $\theta$ & 0.2 & 0.1 & 0.2 & 0.2 & 0.2 & 0.2 \\
        forget. & pow & pow & lin & pow & lin & pow \\
        agg. & sum & avg & avg & sum & avg & sum \\
        \hline
        \hline
        \multicolumn{7}{|c|}{Highschool} \\
        \hline
          & CNS & NSCV & CNSCV & CNSTV & NSCTV & CNSCTV \\
        \hline
        Interval & - & 5 & 5 & 5 & 60 & 60 \\
        $L$ & 72 & 2 & 1.5 & 72 & 1 & 72 \\
        $\mu$ & 0.3 & 0.3 & 0.3 & 0.3 & 0.8 & 0.3 \\
        $\theta$ & 0.1 & 0.1 & 0.2 & 0.2 & 0.2 & 0.2 \\
        forget. & lin & pow & pow & lin & lin & lin \\
        agg. & sum & avg & avg & sum & sum & avg \\
        \hline
        \hline
    \end{tabular}
    \label{tab:appendix_best_param_prec_event}
\end{table}

\begin{table}[htbp]
    \centering
    \caption{The best CogSNet parameters for each method and dataset for precision metric and future link sampling.}
    \begin{tabular}{|c|c|c|c|c|c|c|}
        \hline
        \multicolumn{7}{|c|}{PRECISION - FUTURE LINK SAMPLING} \\
        \hline
        \hline
        \multicolumn{7}{|c|}{Hypertext} \\
        \hline
          & CNS & NSCV & CNSCV & CNSTV & NSCTV & CNSCTV \\
        \hline
        Interval & - & 360 & 360 & 30 & 30 & 30 \\
        $L$ & 48 & 12 & 12 & 6 & 24 & 3 \\
        $\mu$ & 0.3 & 0.8 & 0.5 & 0.8 & 0.8 & 0.8 \\
        $\theta$ & 0.2 & 0.2 & 0.1 & 0.1 & 0.1 & 0.2 \\
        forget. & lin & exp & exp & pow & lin & exp \\
        agg. & sum & avg & avg & sum & sum & sum \\
        \hline
        \hline
        \multicolumn{7}{|c|}{Manufacturing} \\
        \hline
          & CNS & NSCV & CNSCV & CNSTV & NSCTV & CNSCTV \\
        \hline
        Interval & - & 43200 & 43200 & 10080 & 10080 & 10080 \\
        $L$ & 24 & 24 & 24 & 72 & 4320 & 4320 \\
        $\mu$ & 0.5 & 0.5 & 0.3 & 0.3 & 0.8 & 0.3 \\
        $\theta$ & 0.1 & 0.2 & 0.1 & 0.2 & 0.1 & 0.2 \\
        forget. & pow & exp & exp & exp & lin & exp \\
        agg. & sum & avg & avg & sum & sum & sum \\
        \hline
        \hline
        \multicolumn{7}{|c|}{EU-emails} \\
        \hline
          & CNS & NSCV & CNSCV & CNSTV & NSCTV & CNSCTV \\
        \hline
        Interval & - & 1440 & 1440 & 10080 & 1440 & 1440 \\
        $L$ & 168 & 168 & 720 & 168 & 72 & 336 \\
        $\mu$ & 0.3 & 0.5 & 0.5 & 0.3 & 0.3 & 0.3 \\
        $\theta$ & 0.2 & 0.1 & 0.2 & 0.2 & 0.1 & 0.1 \\
        forget. & lin & lin & lin & lin & lin & lin \\
        agg. & sum & avg & avg & sum & sum & sum \\
        \hline
        \hline
        \multicolumn{7}{|c|}{Office} \\
        \hline
          & CNS & NSCV & CNSCV & CNSTV & NSCTV & CNSCTV \\
        \hline
        Interval & - & 5 & 5 & 5 & 120 & 120 \\
        $L$ & 1.5 & 1.5 & 1.5 & 1.5 & 6 & 1.5 \\
        $\mu$ & 0.3 & 0.3 & 0.3 & 0.3 & 0.8 & 0.3 \\
        $\theta$ & 0.2 & 0.2 & 0.2 & 0.2 & 0.1 & 0.2 \\
        forget. & pow & pow & pow & pow & exp & pow \\
        agg. & sum & avg & avg & sum & sum & sum \\
        \hline
        \hline
        \multicolumn{7}{|c|}{Highschool} \\
        \hline
          & CNS & NSCV & CNSCV & CNSTV & NSCTV & CNSCTV \\
        \hline
        Interval & - & 720 & 720 & 5 & 120 & 240 \\
        $L$ & 2 & 12 & 2 & 1.5 & 72 & 72 \\
        $\mu$ & 0.3 & 0.8 & 0.3 & 0.3 & 0.8 & 0.3 \\
        $\theta$ & 0.2 & 0.1 & 0.2 & 0.2 & 0.2 & 0.2 \\
        forget. & pow & exp & pow & pow & exp & lin \\
        agg. & sum & avg & avg & sum & sum & sum \\
        \hline
        \hline
    \end{tabular}
    \label{tab:appendix_best_param_prec_future}
\end{table}

\section{Additional results}
\begin{figure*}[ht]
    \centering
    
    \subfloat[]{\includegraphics[width=0.82\linewidth]{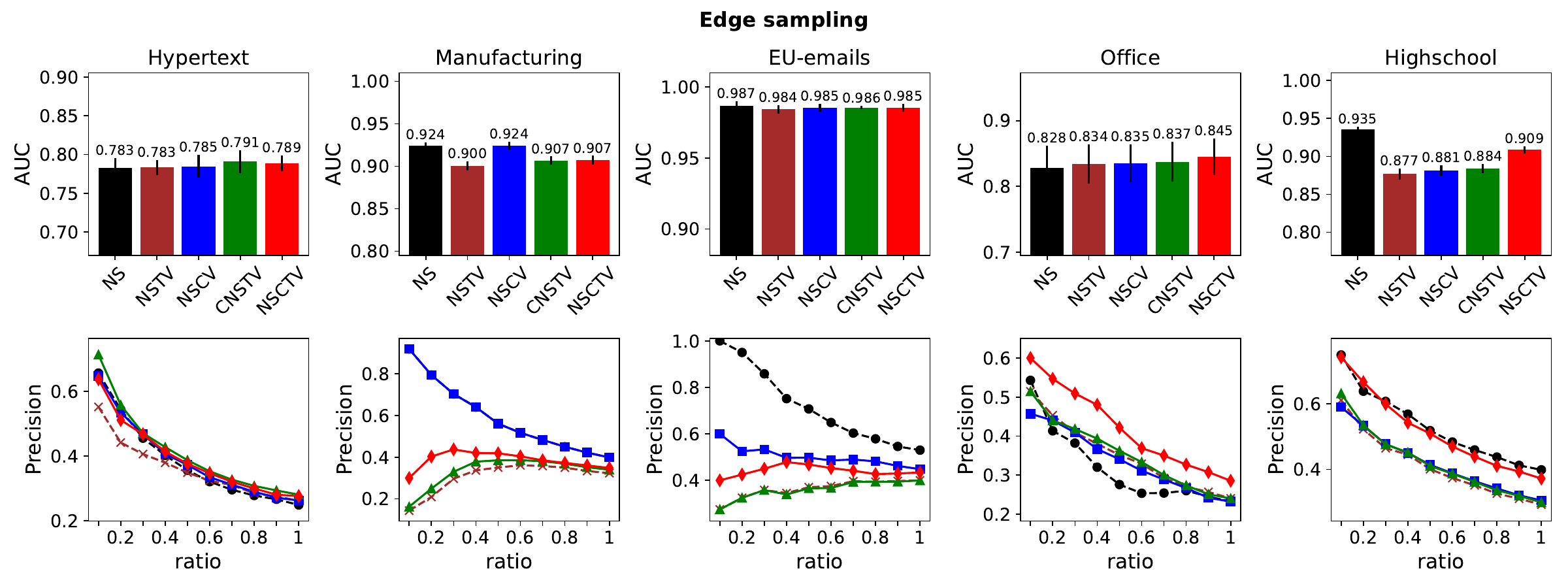}}
    
    \subfloat[]{\includegraphics[width=0.82\linewidth]{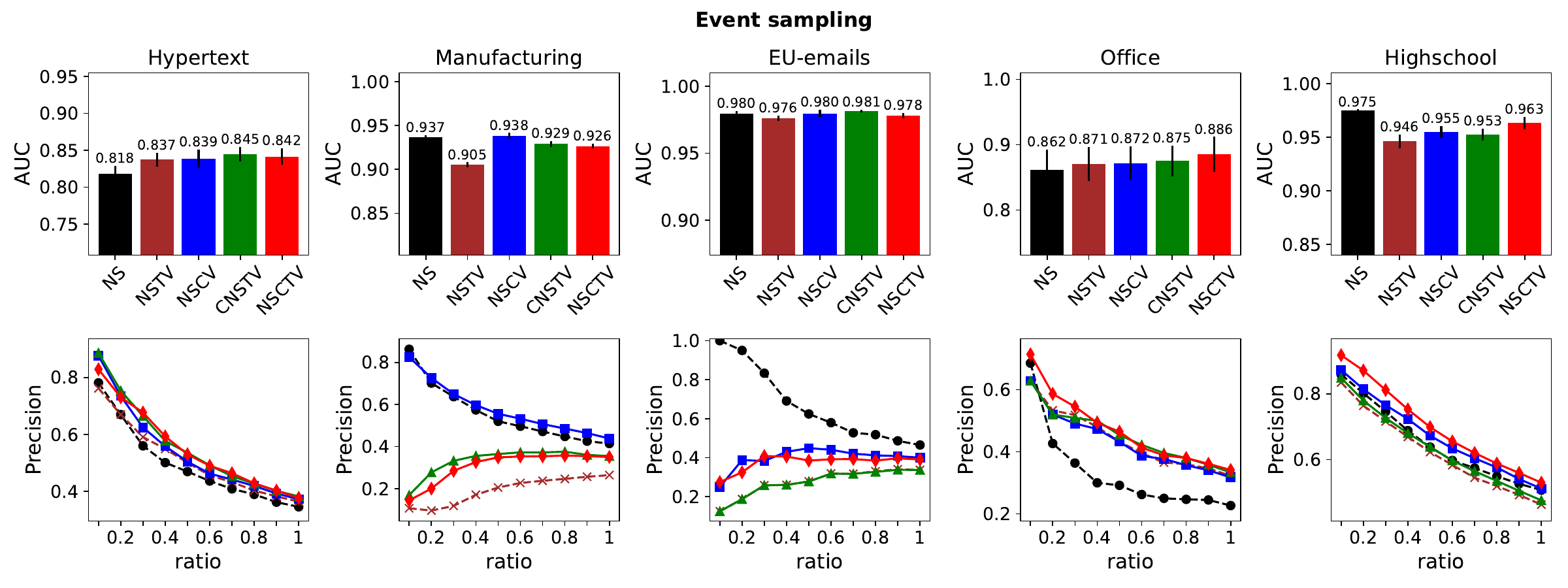}}

    \subfloat[]{\includegraphics[width=0.82\linewidth]{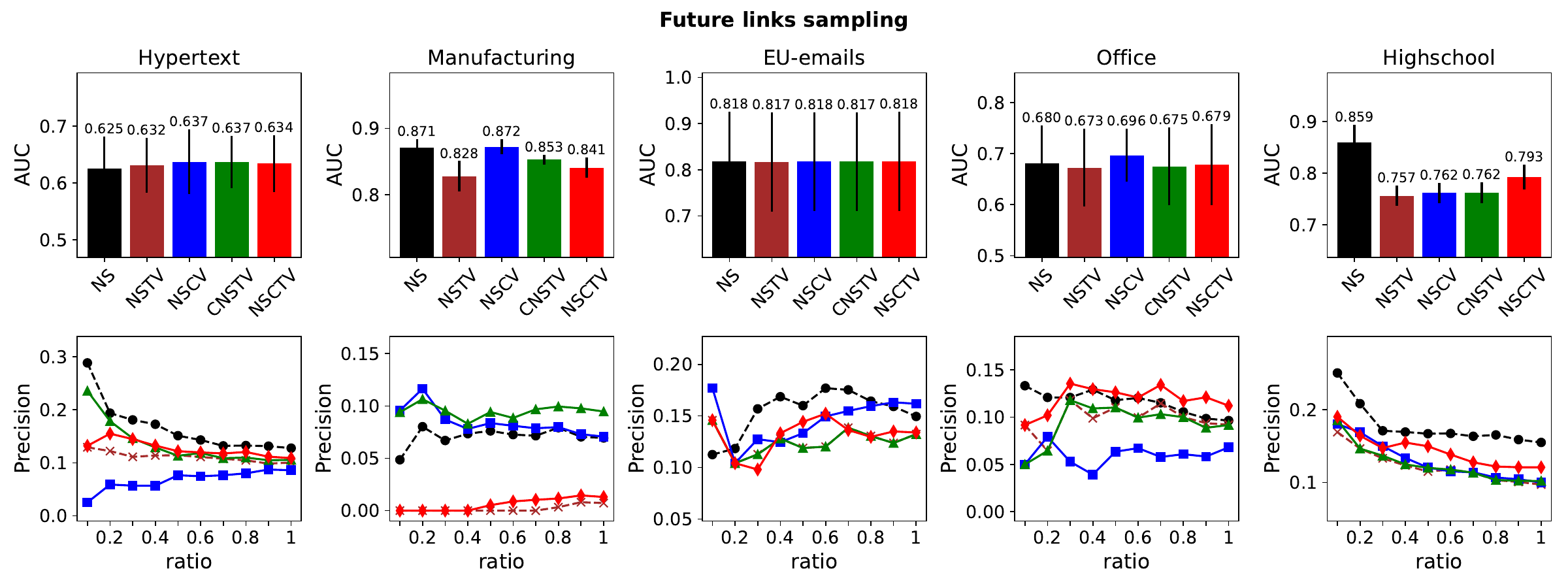}}
    
    \caption{Predictive evaluation of methods based on CogSNet when parameters were selected to maximize AUC for (a) edge sampling, (b) event sampling, and (c) future links sampling. The ratio in precision plots determines the complexity of the problem - a higher value indicates more links to predict. The precision plot uses the same colors for methods as the AUC plot.}
    \label{fig:appendix_auc}
\end{figure*}

\begin{figure*}[ht]
    \centering
    
    \subfloat[]{\includegraphics[width=0.82\linewidth]{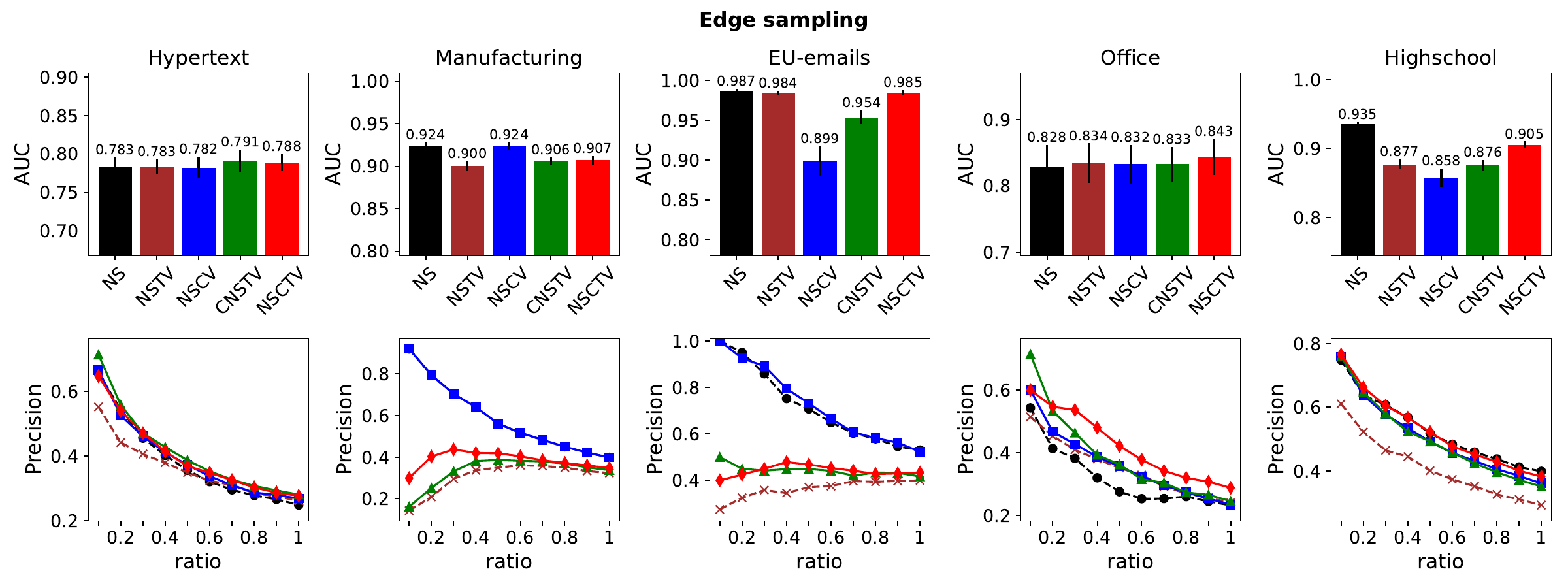}}
    
    \subfloat[]{\includegraphics[width=0.82\linewidth]{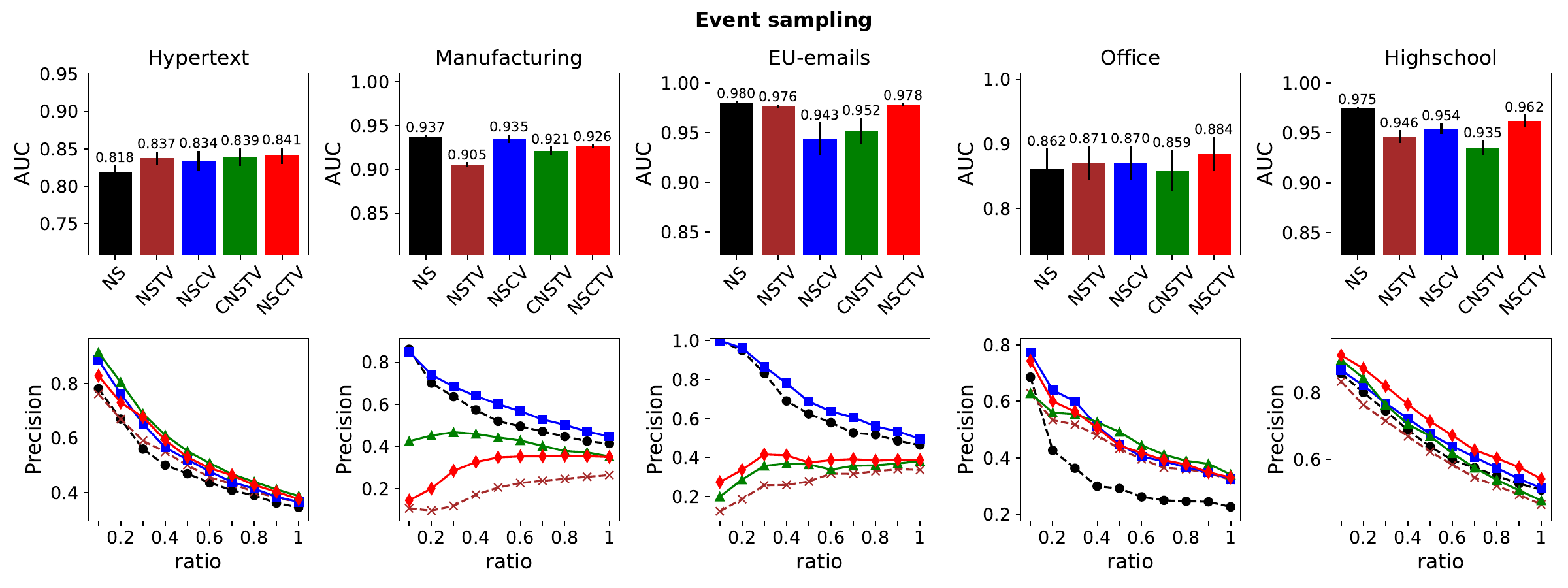}}

    \subfloat[]{\includegraphics[width=0.82\linewidth]{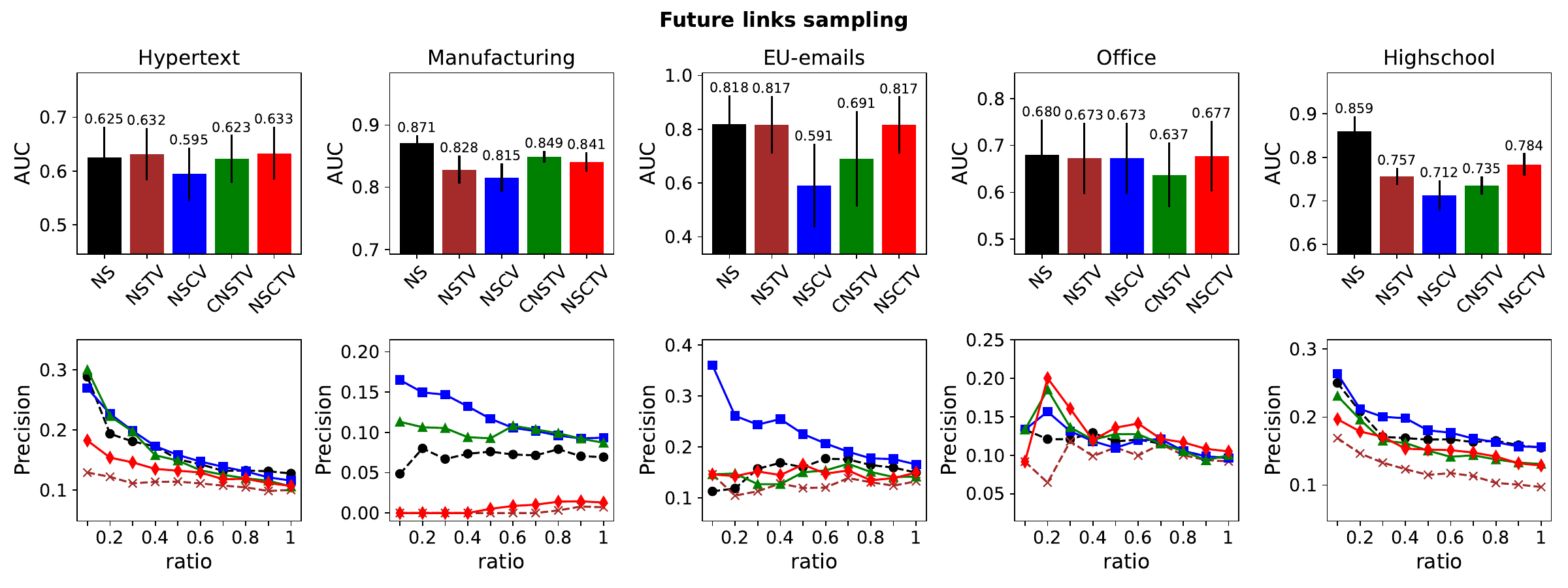}}
    
    \caption{
Predictive evaluation of methods based on CogSNet when parameters were selected to maximize precision for (a) edge sampling, (b) event sampling, and (c) future links sampling. The ratio in precision plots determines the complexity of the problem - a higher value indicates more links to predict. The precision plot uses the same colors for methods as the AUC plot.}
    \label{fig:appendix_prec}
\end{figure*}

The results for the remaining CogSNet-based methods NSCV, CNSTV, and NSCTV. Fig.~\ref{fig:appendix_auc} contains results for parameters selected based on AUC and Fig.~\ref{fig:appendix_prec} contains results for parameters selected based on precision.

\end{document}